\documentclass[aps,pra,showpacs,reprint,superscriptaddress,floatfix]{revtex4-1}
\usepackage{graphicx}
\usepackage{longtable}
\usepackage{dcolumn}
\usepackage{multirow}
\usepackage{amssymb, amsmath}
\usepackage[colorlinks=true, citecolor=blue,linkcolor=blue]{hyperref}
\begin{document}
\title{Calculations of electronic properties and vibrational parameters of alkaline-earth lithides: MgLi$^+$ and CaLi$^+$}
\author{Renu Bala}
\email[Electronic address: ]{rbala@ph.iitr.ac.in}
\author{H. S. Nataraj}
\affiliation{ Department of Physics, Indian Institute of Technology Roorkee,
Roorkee - 247667, India}
\author{Minori Abe}
\affiliation{Department of Chemistry, Tokyo Metropolitan University, 1-1 Minami-Osawa, Hachioji, Tokyo 192-0397, Japan}
\author{Masatoshi Kajita}
\affiliation{National Institute of Information and Communication Technology, Nukui-Kitamachi, Koganei, Tokyo 184-795, Japan}
\begin{abstract}
The $^1\Sigma^+$ electronic ground states of MgLi$^+$ and CaLi$^+$ molecular ions are investigated for their spectroscopic constants and properties such as the dipole- and quadrupole moments, and static dipole polarizabilities. The quadrupole moments and the static dipole polarizabilities for these ions have been calculated and reported here, for the first time. The maximum possible error bars, arising due to the finite basis set and the exclusion of higher correlation effects beyond partial triples, are quoted for reliability. Further, the adiabatic effects such as diagonal Born-Oppenheimer corrections are also calculated for these molecules. The vibrational energies, the wavefunctions, and the relevant vibrational parameters are obtained by solving the vibrational Schr\"{o}dinger equation using the potential energy curve and the permanent dipole moment curve of the molecular electronic ground state. Thereafter, spontaneous and black-body radiation induced transition rates are calculated to obtain the lifetimes of the vibrational states. The lifetime of rovibronic ground state for MgLi$^+$, at room temperature, is found to be $2.81\,\mathrm s$ and for CaLi$^+$ it is $3.19\,\mathrm s$. It has been observed that the lifetime of the highly excited vibrational state is several times larger than (comparable to) that of the vibrational ground state of MgLi$^+$ (CaLi$^+$). In addition, a few  low-lying electronic excited states of  $\Sigma$ and $\Pi$ symmetries have been investigated for their electronic and vibrational properties, using EOM-CCSD method together with the QZ basis sets.\\

\end{abstract}
\maketitle
%
\section{Introduction}
%
The diatomic molecules containing alkaline-earth elements have been studied extensively for over decades and yet, they continue to attract the attention of both experimentalists and theorists alike even today. Some of the alkaline-earth-monofluorides (XF) have been laser cooled and trapped ~\cite{Yin, Tarbutt, Zhelyazkova, Barry, McCarron, Bu, Shuman} for various applications such as for the study of fundamental symmetry violating effects~\cite{ Isaev, Nayak, Kudashov, Anastasia_Borschevsky, Hou, Kozlov}. The alkaline-earth-monohydrides (both XH and XH$^+$) have been observed in several astrophysical atmospheres such as solar, stellar, cometary and interstellar medium \cite{Sinha, Ramachandran, Kirkpatrick, Wallace, Sotirovski}. They have been studied both theoretically~\cite{Abe, Roos, Habli1, Machado, Aymar, Gao1, Mejrissi} and experimentally~\cite{Molhave, Weinstein, Shayesteh}. The BeH molecules are considered for the molecular diagnostics of the fusion plasmas such as the one used in International Thermonuclear Fusion Reactor, ITER \cite{Laporta, Roos, Celiberto}. Further, MgH$^+$, CaH$^+$, and SrH$^+$ have been proposed for the tests of temporal variation in the fundamental physical constants such as the ratio of proton to electron mass (m$_p$/m$_e$) \cite{Kajita2}.\\
The diatomic molecules containing alkali-alkaline-earth-metals (XY; Y = Li, Na, K, Rb, Cs) have strong long-range interactions owing to their large dipole moments \cite{Gopakumar1, Habli}. Their spectroscopic constants and molecular properties have been investigated by several research groups ~\cite{Gopakumar1, Gopakumar2, Ghanmi, Jellali, Rakshit}. Both CaLi and SrLi molecules have also been proposed for the study of m$_p$/m$_e$~\cite{Kajita1}. 
To observe the ultracold atom-ion interactions, the elastic collisions between laser-cooled fermionic lithium atoms and calcium ions have been studied by Haze~\emph{et al.}~\cite{Haze}. The long lifetimes of the highly excited vibrational states in LiBe$^+$, LiMg$^+$, NaBe$^+$, and NaMg$^+$ enable these molecules for being used in ultracold experiments \cite{Fedorov}. 
The aforementioned reasons call for the accurate theoretical data on the molecules that are considered in this work. This may further be used for predicting the formation of such molecules and to control their internal and external degrees of freedom.\\
Although there exists some theoretical studies for MgLi$^+$ and CaLi$^+$, there is no experimental data available in the literature yet. The spectroscopic constants such as bond length ($R_e$), dissociation energy ($D_e$), and harmonic frequency ($\omega_e$) are calculated for MgLi$^+$ using the primitive 6-31G$^*$ basis set together with Hartree-Fock (HF) and M{\o}ller-Plesset perturbation method of second-order (MP2)  by Pyykk\"o~\emph{et al.}~\cite{Pyykko}. For the same ion, Boldyrev~\emph{et al.}~\cite{Boldyrev} have calculated $R_e$, $\omega_e$ at MP2 (full) level, and $D_e$ at many-body perturbation theory of fourth-order (MP4), and quadratic configuration interaction with singles and doubles including perturbative triples (QCISD(T)) level, using 6-311+G$^*$ basis set for the ground state. An \emph{ab initio} study of ground and low lying excited states of MgLi and MgLi$^+$ have been reported by Gao and Gao~\cite{Gao} using valence full configuration interaction (FCI) and multi-reference configuration interaction (MRCI) method. The study of spectroscopic constants, potential energy curves (PECs), permanent and transition electric dipole moment curves of ground - and several excited states of MgLi$^+$ have been reported by ElOualhazi and Berriche~\cite{ElOualhazi} using CIPSI package (configuration interaction by perturbation of a multiconfiguration wavefunction selected iteratively). They have used nonempirical pseudopotential approach that considers the MgLi$^+$ molecular ion as an effective two electron system moving in the effective potential created by Mg$^{2+}$ and Li$^+$ core. \emph{Ab initio} calculations for ground state spectroscopic constants, permanent dipole moments (PDMs) and lifetimes of alkali-alkaline-earth cations have been reported by Fedorov~\emph{et al.}~\cite{Fedorov}. \\ 
The perturbed-stationary-state method has been employed by Kimura~\emph{et al.}~\cite{Kimura} to charge transfer in Li$^+$\,+\,Ca collisions and have calculated the diatomic constants for the ground state using pseudo-potential technique and Slater-type orbital basis sets. Russon~\emph{et al.}~\cite{Russon} have reported \emph{ab initio} all-electron calculations for $R_e$, $\omega_e$ and ionization energy of X\,$^1\Sigma^+$ state using QCISD(T) and also using complete active space self-consistent-field MRCI with single- and double-excitation (CASSCF-MRCISD) method. Habli~\emph{et al.}~\cite{Habli} also have performed full CI calculations for electronic and vibrational properties of ground - and excited states of CaLi$^+$ by treating two valence electrons explicitly with core polarization potentials (CPP) approach.\\
In the current work, we have performed the calculations of PECs, the spectroscopic constants: $R_e$, $D_e$, $\omega_e$, $\omega_ex_e$, $B_e$ and $\alpha_e$, and the molecular properties: dipole moment ($\mu_z$), quadrupole moment ($\Theta_{zz}$), average or isotropic polarizability ($\bar{\alpha}$), polarizability anisotropy ($\gamma$) at molecular equilibrium point, and the $z$-component of polarizability at the super-molecular limit, for the ground state of MgLi$^+$ and CaLi$^+$ ionic systems at different levels of correlation with fully optimized basis sets with an objective of computing the reported results for both the molecular ions at the same level of accuracy. 
Further, vibrational spectroscopic calculations are performed by solving vibrational Schr\"odinger equation using PECs and PDM curves. Furthermore, we have extended this work to the lower excited states for both the molecular ions using equation-of-motion CCSD (EOM-CCSD) method. The current work is a sequel to our previous paper on BeLi$^+$~\cite{BeLi+}.\\ 
The paper is presented in three other sections: the introduction is followed by a brief description of the methods involved in Section \ref{section-2}, the detailed description of the results in Section \ref{section-3} and finally the summary of the present work in Section \ref{section-4}.
%
\begin{table*}[ht]
\begin{ruledtabular}
\begin{center}
\caption{\label{table-I} The spectroscopic constants for the electronic ground state of MgLi$^+$ and CaLi$^+$  at various levels of correlation, computed for the non-relativistic case, and compared with the published results.}
\begin{tabular}{ccccccccc}
Molecule & Method/basis & $R_e$ (a.u.) & $D_e$ (cm$^{-1}$) & $\omega$$_e$(cm$^{-1}$) & $\omega$$_ex_e$(cm$^{-1}$) & $B_e$ (cm$^{-1}$) & $\alpha_e$ (cm$^{-1}$) \\
  \hline 
MgLi$^+$ & SCF/TZ & 5.599 & 6396.1 & 256.8 & 2.47 & 0.3537 & 0.0048   \\
         & MP2/TZ & 5.516 & 6595.9 & 264.5 & 2.46 & 0.3645 &  0.0049\\
         & CCSD/TZ & 5.510 & 6634.6 & 265.8 & 2.49 & 0.3652  & 0.0047\\
         & CCSD(T)/TZ& 5.509 & 6641.9 & 265.7 & 2.26 & 0.3655 & 0.0047 \\
         \hline
     & SCF/QZ & 5.599 & 6409.3 & 256.6 & 2.49 & 0.3537 & 0.0049 \\
     & MP2/QZ & 5.501 & 6659.7 & 265.8 & 2.49 & 0.3665 & 0.0048\\
     & CCSD/QZ & 5.497 & 6702.2 & 267.0 & 2.32 & 0.3670 & 0.0047 \\
     & \textbf{CCSD(T)/QZ} & \textbf{5.493} & \textbf{6712.4} & \textbf{267.3} & \textbf{2.30} & \textbf{0.3675} & \textbf{0.0046} \\ 
     & \textbf{Error bar} & \textbf{$\pm$0.02} & \textbf{$\pm$80.7} & \textbf{$\pm$1.9} & \textbf{$\pm$0.06} & \textbf{$\pm$0.0025} & \textbf{$\pm$0.0002}\\
     \hline
          & CCSDT/cc-pCVQZ~\cite{Fedorov} & 5.476 & 6658.8 & 265.9 & 2.0 & $-$ & $-$\\
          & MRCI/cc-pCVQZ~\cite{Fedorov} & 5.481 & 6649.2 & 265.4 & 2.0 & $-$ & $-$\\
     & HF/6-31 G$^*$~\cite{Pyykko} & 5.629 & 6264.5 & 255 & $-$ & $-$ & $-$\\
     & MP2/6-31 G$^*$~\cite{Pyykko}& 5.569 & 6456.4 & 262 & $-$ & $-$ & $-$\\
     & MP2(full)/6-311 +G$^*$~\cite{Boldyrev} & 5.546 & 6470.5 & 261 & $-$ & $-$ & $-$ \\
     & MRCI/AV5Z+Q~\cite{Gao} & 5.546 & 6508.9 & 263.5 & 2.37 & 0.3606 & $-$\\ 
     & MRCI/AV5Z+Q+DK~\cite{Gao} & 5.533 & 6557.3 & 266.4 & 2.48 & 0.3623  & $-$\\ 
     & MRCI/AVQZ+Q~\cite{Gao}  & 5.548 & 6484.7 & 262.9 & 2.35 & 0.3603 & $-$\\
     & MRCI/AVQZ+Q+DK~\cite{Gao}  & 5.544 & 6476.6 & 262.9 & 2.36 & 0.3608 & $-$\\
     & FCI/Gaussian~\cite{ElOualhazi} & 5.47 & 6575 & 264.22 & 2.63 & 0.372138 & $-$\\
     & SCF/STO-6G ~\cite{Fantucci}  & 5.633 & $-$ & $-$ & $-$ & $-$  & $-$\\
     \hline \hline    
CaLi$^+$ & SCF/TZ & 6.329 & 9734.4 & 240.2 & 1.79 & 0.2518 & 0.0024\\
          & MP2/TZ & 6.178 & 9328.0 & 246.6 & 1.82 & 0.2643 & 0.0026\\
          & CCSD/TZ & 6.189 & 9987.6 & 247.6 & 1.43 & 0.2633 & 0.0024\\
          & CCSD(T)/TZ& 6.193 & 10008.5 & 245.8 & 1.60 & 0.2630 & 0.0024\\
          \hline
          & SCF/QZ & 6.327 & 9744.7 & 240.3 & 1.67 & 0.2520 & 0.0024\\
          & MP2/QZ & 6.153 & 9428.0 & 246.2 & 1.68 & 0.2664 & 0.0025\\
          & CCSD/QZ & 6.172 &  10083.4 &  246.8 & 1.37 & 0.2648 & 0.0023\\
          & \textbf{CCSD(T)/QZ} & \textbf{6.170} & \textbf{10092.9} & \textbf{245.1} & \textbf{1.31} & \textbf{0.2650} &  \textbf{0.0023}\\
          & \textbf{Error bar }  &  \textbf{$\pm$0.025} &  \textbf{$\pm$94} &  \textbf{$\pm$2.4} &  \textbf{$\pm$0.23} &  \textbf{$\pm$0.0022} & \textbf{$\pm$0.0001 }\\
          \hline
          & FCI/Gaussian ~\cite{Habli} & 6.120 & 9973.27 & 242 & $-$ & $-$ & $-$\\
          & FCI/STO's ~\cite{Kimura} & 6.20 & 8952.8 & 235 & $-$ & 0.263 & $-$ \\
          & QCISD(T) ~\cite{Russon}  & 6.274 & 10012.7 & 239 & $-$ & $-$ & $-$\\
          & CASSCF-MRCISD ~\cite{Russon} & 6.272 & $-$ & 239 & $-$ & $-$ & $-$ \\
          & QCISD(T,full) ~\cite{Russon} & $-$ & 9678.7 & $-$ & $-$ & $-$ & $-$\\
\end{tabular}
\begin{flushleft}
\end{flushleft}
\end{center}
\end{ruledtabular}
\end{table*}
%
\section{Methodology}
\label{section-2}
%
The electronic structure calculations reported in this work are carried out systematically at different levels of correlation: self-consistent field (SCF), many-body perturbation theory (MP2), coupled-cluster method with single and double excitations (CCSD) and CCSD with partial triples (CCSD(T)) using CFOUR~\cite{CFOUR} and DIRAC15~\cite{DIRAC} software packages, for non-relativistic and relativistic cases, respectively. The nuclear masses used for Li, Mg and Ca are $7.01600$ a.u., $23.98504$ a.u., and $39.96259$ a.u., respectively. The uncontracted correlation-consistent polarized valence (cc-pV) triple-zeta (TZ) and quadruple-zeta (QZ) basis sets are used~\cite{Dunning, Koput} together with the C$_{2v}$ point group symmetry. Despite using such large basis sets, neither the valence orbitals are truncated nor the core electrons are frozen. All the electrons, {\itshape viz.} core, valence, and virtual orbitals, are kept active for the calculations of diatomic constants, and molecular properties that are reported in this paper. 
The center-of-mass of the molecular ion is taken as the origin of the coordinate system. Further, the electronic energies are computed at different internuclear distances ($R$) ranging from $0.4${\,\AA} to $30${\,\AA} with a step size of $1${\,\AA}. Also, around the vicinity of the potential minimum, a small refined step size of $0.001${\,\AA} is considered.  From the PECs that are plotted, the dissociation energies are calculated as a difference between the energies at equilibrium point and those at a distance of 30{\,\AA}. Note that the energy difference between this ceiling distance and that of conventional  asymptotic limit of $52${\,\AA} or $100$ a.u. distance is less than $0.7$ cm$^{-1}$ for MgLi$^+$ and $1.5$ cm$^{-1}$ for CaLi$^+$. \\ 
We have calculated the spectroscopic constants, and molecular properties for the electronic ground state at non-relativistic level using CFOUR program. The basis sets have been taken from the EMSL library~\cite{EMSL}. The harmonic frequencies and anharmonic constants are calculated using second-order vibrational perturbation theory adopted in the CFOUR package.
\begin{figure}[h]
\includegraphics[clip,width=\columnwidth]{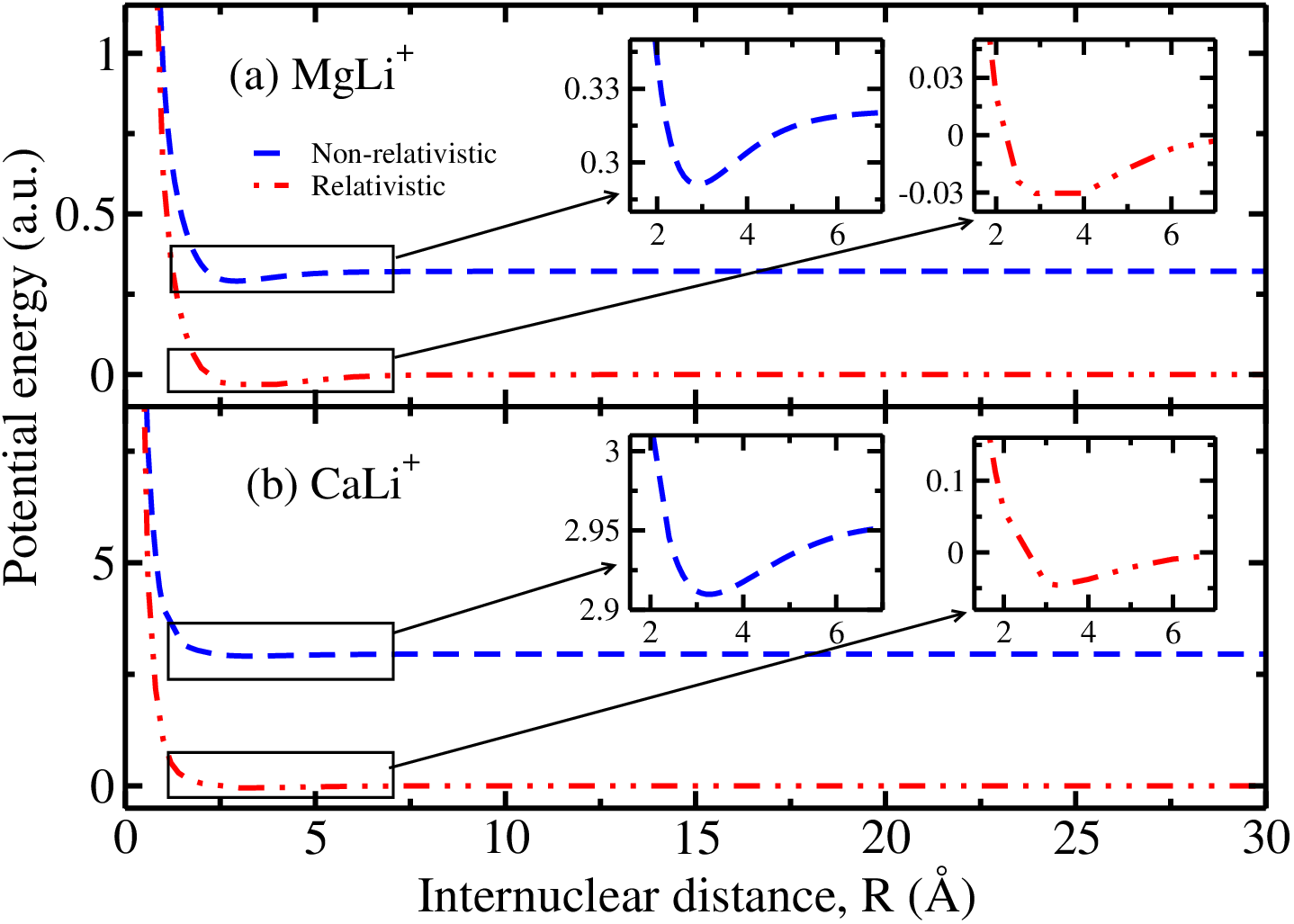}
\caption{\label{fig1}(colour online) Non-relativistic and relativistic potential energy curves plotted with respect to the relativistic dissociation energy calculated using CCSD(T)/QZ method.}
\end{figure}
%
By convention, the internuclear axis is taken to be along the \textit{z}-direction and hence, $\alpha_{zz}\,\equiv\,\alpha{_\parallel}$. Similarly, on calculating the other two perpendicular components ($\alpha{_\bot}$) of dipole polarizability (i.e., $\alpha_{xx}$ and $\alpha_{yy}$), we have obtained the average polarizability, $\bar{\alpha}$, and the anisotropic polarizability, $\gamma$ as,
  \begin{eqnarray}{}
\bar{\alpha}\,=\,(\alpha{_\parallel }+2\alpha{_\bot})/3,\quad  \mathrm{and} \quad \gamma\,=\,\alpha{_\parallel}-\alpha{_\bot}.
\end{eqnarray}
The components of the traceless ($\sum_i\Theta_{i\,i}=0$) quadrupole moment tensor are given by,
\begin{eqnarray}{}
\Theta_{i\,j}\,=\,\frac{q}{2}\sum\limits_{i,\,j} \left(3\,r_{i}\,r_{j}\,-\,r^2\,\delta_{i\,j}\right).
\end{eqnarray}
Further, for diatomic molecules we have $\Theta_{xx}\,=\,\Theta_{yy}$, and hence, we can write,  
\begin{eqnarray}{}
   \Theta_{zz}\,=\,-\,2\,\Theta_{xx}.
\end{eqnarray}
Using the PECs shown in Figure~\ref{fig1} and PDM curves shown in Figure~\ref{fig2}, 
we have solved the vibrational Schr\"odinger equation using Le Roy's LEVEL16 program~\cite{LEVEL} and obtained the vibrational parameters {\itshape viz.} wavefunctions, energy levels ($E_v$), transition dipole moments (TDMs) between different vibrational states, and vibrationally coupled rotational constants ($B_v$). An appropriate step size required for numerical calculations of these parameters is estimated using equation ($3$) of Ref.~\cite{LEVEL} and cubic spline fitting is used for interpolation.   
Further, the vibrational energy differences and TDM values between the vibrational states are used to calculate the spontaneous and black-body radiation- (BBR-) induced transition rates at the surrounding temperature, {\itshape viz.} $T\,=\,300$\,K as~\cite{Kotochigova}, 
\begin{eqnarray}{}  
\Gamma_{v, J}^{spon}\,&=&\,\sum\limits_{v^{''}, J^{''}}\Gamma^{emis}(v, J\,\rightarrow\,v^{''}, J^{''}),\\ 
 \Gamma_{v, J}^{BBR}\,&=& \,\sum\limits_{v^{''}, J^{''}}\bar{n}(\omega)\,\Gamma^{emis}(v, J\,\rightarrow\,v^{''}, J^{''})\nonumber\\
 &+&\sum\limits_{v^{'}, J^{'}}\bar{n}(\omega)\,\Gamma^{abs}(v, J\,\rightarrow\,v^{'}, J^{'}),
  \end{eqnarray}
where the indices ($v^{''}, J^{''}$) and ($v^{'}, J^{'}$) denote the rovibrational states, within the same electronic state, whose energies are, respectively, lower and higher than that of ($v, J$) level.  
The average number of photons $\bar{n}(\omega)$ in a single mode at frequency $\omega$ is given by the relation, \\
  \begin{eqnarray}{}
  \bar{n}(\omega)\,\,=\,\,\frac{1}{exp{(\hbar\omega/k_{B}T)}-1},
   \end{eqnarray}
where $\hbar\,\omega\,=\,\arrowvert E_{v, J}-E_{\tilde{v}, \tilde{J}}\arrowvert$ is the energy difference between the two rovibrational levels involved, where ($\tilde v, \tilde J$) is ($v^{''}, J^{''}$) for emission, while ($v^{'}, J^{'}$) for absorption, and $k_{B}$ is the Boltzmann constant. The emission and absorption rates are calculated using the equation,
 \begin{eqnarray}{}
 \Gamma (v, J\,\rightarrow\,\tilde{v}, \tilde{J})\,= \,\frac{8\pi}{3\epsilon_0}\,\frac{1}{h\,c^3}\, \omega^3\, (TDM_{v, J\,\rightarrow\, \tilde{v}, \tilde{J}})^2.
  \end{eqnarray}
Finally, the lifetime ($\tau$) of the rovibrational state is obtained as,
\begin{eqnarray}{}
\tau\,=\,\frac{1}{\Gamma^{total}}; \quad \text {where,\quad} \Gamma^{total}\,=\,\Gamma^{spon}\,+\,\Gamma^{BBR}.
\end{eqnarray}
Further, electronic energies of the low-lying excited states are calculated, by keeping all occupied and virtual orbitals as active, using EOM-CCSD method implemented in CFOUR program taking the QZ basis set. The spectroscopic constants are obtained using VIBROT program available in MOLCAS package~\cite{MOLCAS}. Furthermore, the vibrational parameters for the excited states are computed using LEVEL16 program. \\ 
The relativistic calculations for the ground state molecular constants are carried out using DIRAC$15$ software package. After generating the reference state using SCF method, the energy calculations at MP2, CCSD and CCSD(T) level are carried out using RELCCSD module. The Dirac-Coulomb Hamiltonian is used with the approximation proposed by Visscher~\cite{Visscher} in which contribution from the (SS$|$SS) integrals is replaced by an interatomic (SS$|$SS) correction. The diatomic constants are calculated using VIBROT program. \\
Errors in our calculation may come from two sources: one, from the choice of the basis set - particularly its size, and the other, from the higher-order terms in the CC expansion that are neglected in the present work. The error due to the former source ($\Delta_1$) is not expected to be larger than the difference between the TZ and the QZ results, and yet, we have taken this entire difference as the maximum possible error. Since the contribution from the  correlation effects converges with the higher hierarchy of correlation levels, the error ($\Delta_2$) due to this source is not expected to be larger than the difference, {\itshape viz.}, CCSD(T) - CCSD. With this, the final value of the calculated property ($P$) can be quoted as,
\begin{eqnarray}{}
P_{final}\,=\,P_{CCSD(T)@QZ}\,\pm\arrowvert\Delta_1\arrowvert\,\pm\arrowvert\Delta_2\arrowvert,
\end{eqnarray}
where $P_{CCSD(T)@QZ}$ is the result of the property of interest calculated using QZ basis set at the CCSD(T) level of approximation.\\
%
%
\begin{table*}[ht]
\begin{ruledtabular}
\begin{center}
\caption{\label{table-II} The computed spectroscopic constants for the electronic ground state of MgLi$^+$ and CaLi$^+$ at different levels of correlation, in the relativistic case.}

\begin{tabular}{cccccccc}
 Molecule & Method/basis &$R_e$ (a.u.) & $D_e$ (cm$^{-1}$) & $\omega$$_e$ (cm$^{-1}$) & $\omega$$_ex_e$ (cm$^{-1}$) & $B_e$ (cm$^{-1}$)& $\alpha_e$ (cm$^{-1}$)\\
  \hline 
 MgLi$^+$ & SCF/TZ & 5.597 & 6377.6 & 254 & 2.25 & 0.3484 & 0.0036\\
           & MP2/TZ & 5.513 & 6579.2 & 261.7 & 2.17 & 0.3569 & 0.0031 \\
           & CCSD/TZ & 5.508 & 6618.7 & 264.1 & 2.11 & 0.3580 & 0.0029 \\
           & CCSD(T)/TZ & 5.505 & 6625.9 & 263.4 & 2.03 & 0.3593 & 0.0031 \\
           \hline
          & SCF/QZ & 5.595 & 6391.3 & 254.5 & 2.29 & 0.3495 & 0.0038 \\
          & MP2/QZ & 5.497 & 6644.5 & 263.9 & 2.26 & 0.3597 & 0.0033 \\
          & CCSD/QZ & 5.493 & 6685.7 & 266 & 2.14 & 0.3624 & 0.0034 \\
          & \textbf{CCSD(T)/QZ} & \textbf{5.490} & \textbf{6696.1} & \textbf{265.7} & \textbf{2.12} & \textbf{0.3628} & \textbf{0.0034}\\
          & \textbf{Error bar }  & $\pm$\textbf{0.018} & $\pm$\textbf{80.6} & $\pm$\textbf{2.6} & $\pm$\textbf{0.11} & $\pm$\textbf{0.0039} & $\pm$\textbf{0.0003}\\                  
          \hline \hline                 
 CaLi$^+$ & SCF/TZ & 6.321 & 9649.0 & 240.2 & 1.45 & 0.2475 & 0.0018\\
          &  MP2/TZ & 6.169 & 9323.9 & 246.3 & 1.51 & 0.2586 & 0.0018\\
          & CCSD/TZ & 6.182 & 9911.2 & 245.5 & 1.17 & 0.2554 & 0.0013 \\
          & CCSD(T)/TZ & 6.185 & 9933.8 & 243.3 & 1.08 & 0.2563 & 0.0014 \\
          \hline
          & SCF/QZ & 6.321 & 9661.1 & 240.7 &  1.50& 0.2488 & 0.0020\\
          & MP2/QZ & 6.147 & 9355.4 & 246.4 & 1.52 & 0.2622 & 0.0020\\
          & CCSD/QZ & 6.166 & 9999.9 & 245.2 & 1.18 & 0.2575 & 0.0014\\
          & \textbf{CCSD(T)/QZ} & \textbf{6.165} & \textbf{10010.4} & \textbf{243.2} & \textbf{1.11} & \textbf{0.2561} & \textbf{0.0012}\\ 
          & \textbf{Error bar}  & $\pm$\textbf{0.021} & $\pm$\textbf{87.1} & $\pm$\textbf{2.1} & $\pm$\textbf{0.1} & $\pm$\textbf{0.0016} & $\pm$\textbf{0.0004} \\                   
\end{tabular}
\begin{flushleft}
\end{flushleft}
\end{center}
\end{ruledtabular}
\end{table*}
%
%
\begin{figure}[ht]
\includegraphics[clip,width=\columnwidth]{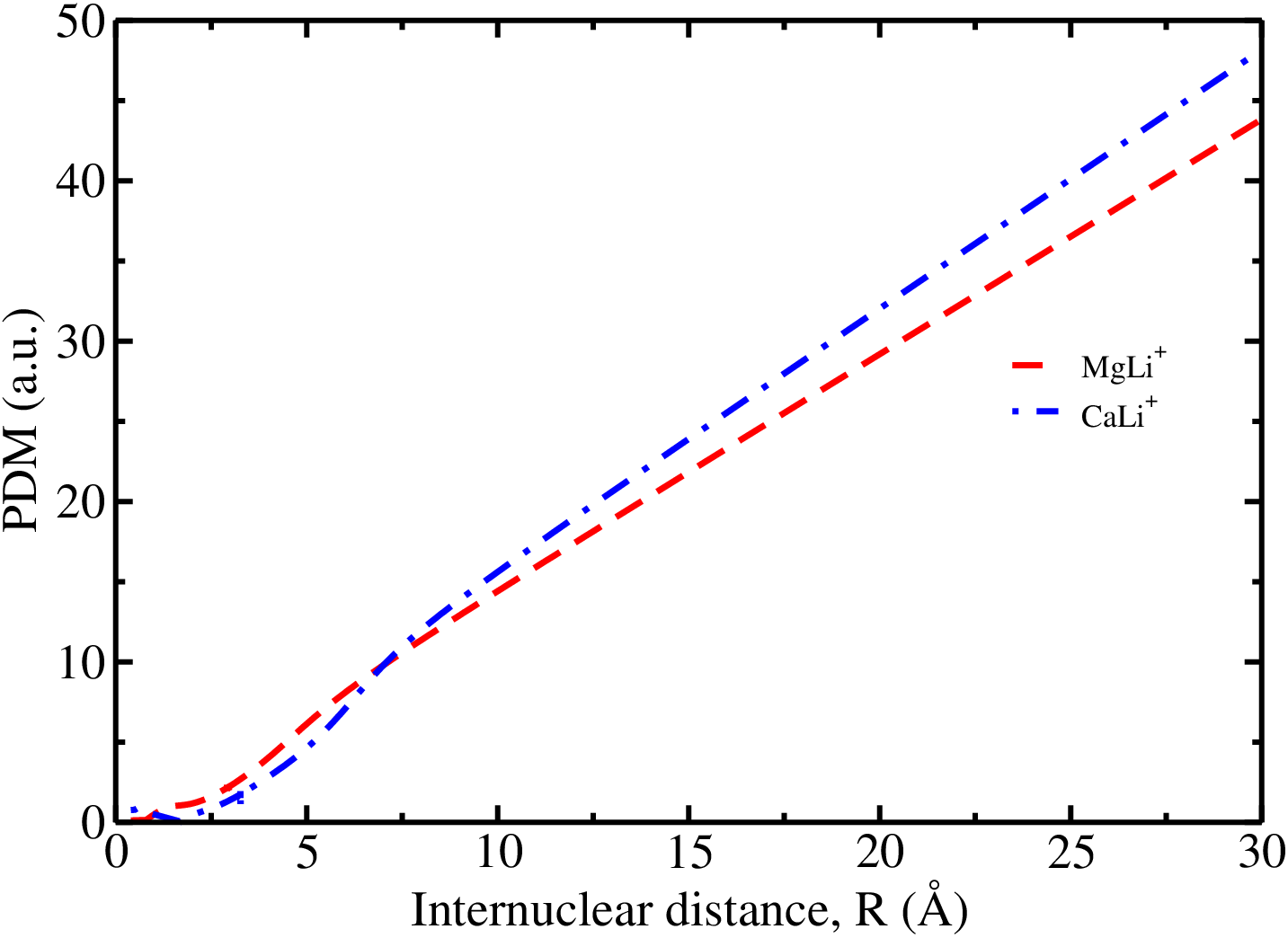}
\caption{\label{fig2}(colour online) Permanent dipole moment curve for the ground state of MgLi$^+$ and CaLi$^+$ using CCSD(T)/QZ method.}
\end{figure}
%
\section{Results and discussion}
\label{section-3}
%
\subsection{Ground state spectroscopic constants and molecular properties}
It has to be noted that the atomic units for distance ($1$\,a.u.\;$=$\;$0.52917721${\,\AA}) and dipole moment ($1$\,a.u.\;$=$\;$2.54174691$\,D) are used, unless otherwise mentioned, throughout the paper. The ground state PECs of MgLi$^+$ and CaLi$^+$ molecular ions are shown in Figure~\ref{fig1}. The dissociation energy calculated in the relativistic case is considered as reference with respect to which the potential energies calculated at different internuclear distances, both in relativistic and non-relativistic cases, are plotted. These PECs reflect the well known fact that the relativistic energy levels are lower than their non-relativistic counterparts. The computed results of the spectroscopic constants at the non-relativistic level are compared with the available results in the literature in Table~\ref{table-I} and their relativistic variants are shown in Table~\ref{table-II}. The relativistic contributions to the spectroscopic constants: $R_e$, $D_e$, and $\omega_e$ are smaller than the error bars quoted in Table~\ref{table-I}. In order to understand whether or not our results of the spectroscopic constants are saturated with respect to basis set size, we have performed CCSD(T) calculations using 5Z basis set for both the ions considered in this work. For MgLi$^+$, the calculated results are: $R_e = 5.452$ a.u., $D_e = 6748.39$ cm$^{-1}$, $\omega_e = 268.8$ cm$^{-1}$, $\omega_ex_e = 2.25$ cm$^{-1}$, $B_e = 0.3731$ cm$^{-1}$ and $\alpha_e = 0.0052$ cm$^{-1}$, while for CaLi$^+$: $R_e = 6.142$ a.u., $D_e = 10134.28$ cm$^{-1}$, $\omega_e = 246.96$ cm$^{-1}$,  $\omega_ex_e = 1.48$ cm$^{-1}$, $B_e = 0.2674$ cm$^{-1}$, and $\alpha_e = 0.0025$ cm$^{-1}$. 
Further, to estimate the higher - order correlation effects beyond CCSD(T)/QZ, we have calculated the diatomic constants using the CCSD with full triples, {\itshape viz.} CCSDT/QZ method. From the comparison of these two sets of results, we infer that the contribution of the missing triples in CCSD(T) method is negligible, at least up to the accuracies that we have reported here, for the diatomic constants $R_e$, $B_e$ and $\alpha_e$ for both the molecular ions. However, $D_e$ increases by $4.91$ cm$^{-1}$ and $27.61$ cm$^{-1}$, while $\omega_e$ decreases by $0.02$ cm$^{-1}$ and $0.24$ cm$^{-1}$, for MgLi$^+$ and CaLi$^+$, respectively. In addition, the value of anharmonic constant changes by -$0.22$ cm$^{-1}$ for MgLi$^+$ and $0.17$ cm$^{-1}$ for CaLi$^+$. Except $\omega_e x_e$ for MgLi$^+$, the contributions of the non-leading order triples to CCSD(T) are well within the error bars that are reported in Table~\ref{table-I}. From these two additional calculations; one using CCSD(T)/5Z and the other using CCSDT/QZ, we infer that our results reported in Table~\ref{table-I} are more saturated with respect to correlation effects than the basis set size effects. For achieving better accuracies, as a trade-off to higher-order correlation effects, one may need to consider either the basis sets larger than 5Z or one may extrapolate the results to complete basis set limit. \\
The diagonal Born-Oppenheimer correction (DBOC) is calculated for both the ions at the CCSD level of theory in conjunction with the QZ basis sets. In the vicinity of the equilibrium geometry, the DBOC is found to be $1001.32$ cm$^{-1}$ and $1823.36$ cm$^{-1}$ for MgLi$^+$ and CaLi$^+$, respectively. Although the magnitude of DBOC is significant, its inclusion to PEC does not seem to affect $R_e$. However, it lowers $D_e$ of MgLi$^+$ by $0.65$ cm$^{-1}$ and $D_e$ of CaLi$^+$  by $1.35$ cm$^{-1}$. Thus, it is very small when compared to the error bars quoted on $D_e$ in Table~\ref{table-I} and Table~\ref{table-II}.\\  
We have calculated several molecular properties such as the electric dipole moments, quadrupole moments and static dipole polarizabilities, however, all at non-relativistic level. These results are tabulated in Table~\ref{table-III}. 
The parallel-component of the dipole polarizability at super-molecular limit, i.e. at a bond distance of $100$\,a.u., denoted as $\alpha_{100}$, is also reported in the last column of Table~\ref{table-III}.
%
\begin{figure}[ht]
\includegraphics[clip,width=\columnwidth]{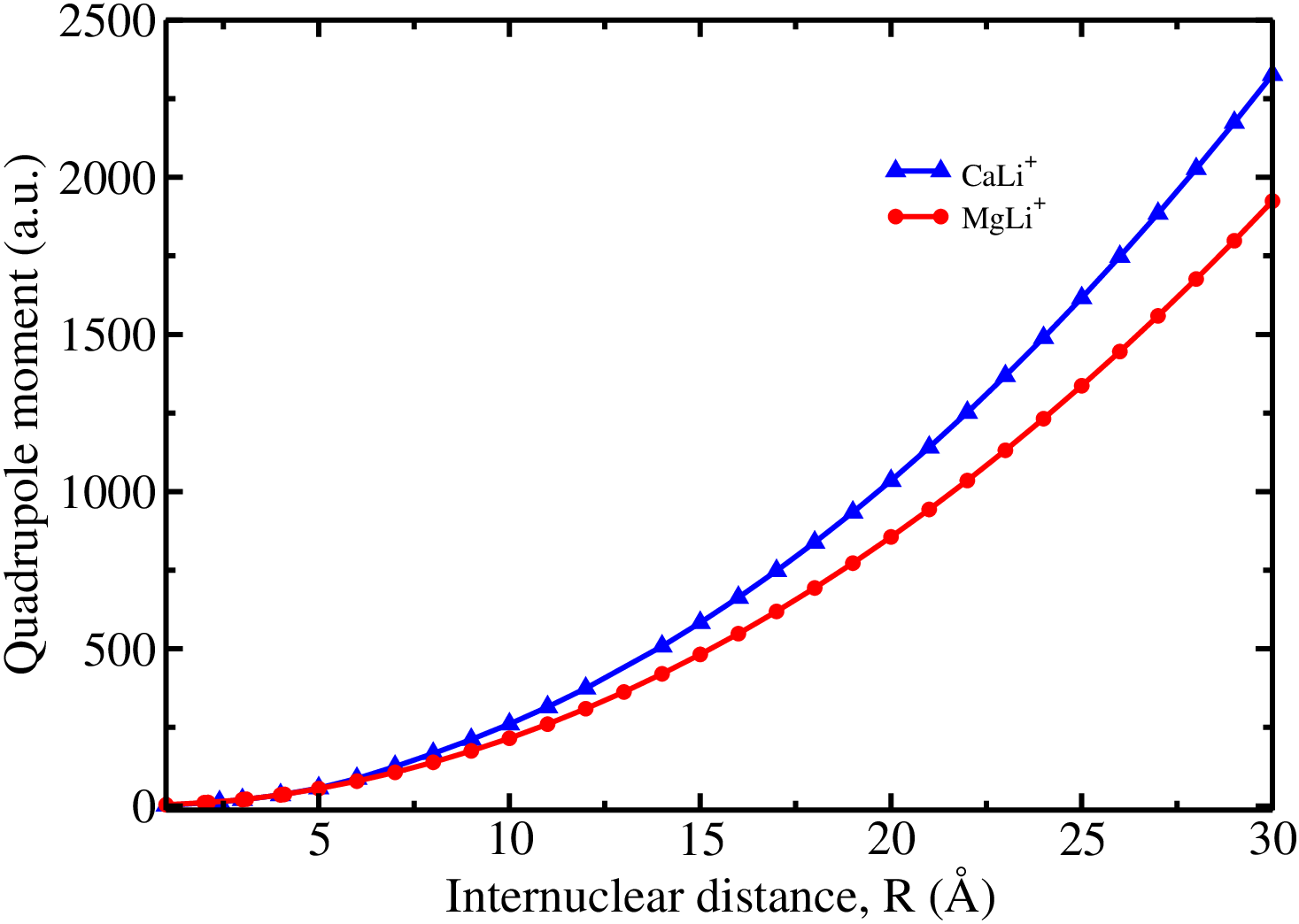}
\caption{\label{fig3}(colour online) Quadrupole moment curve for the ground state of MgLi$^+$ and CaLi$^+$ using CCSD(T)/QZ method.}
\end{figure}
%
With the convention that the direction of interatomic axis is considered from the heavier element to the lighter one, 
we obtain the dipole moment for the ground state of MgLi$^+$ and CaLi$^+$ at the equilibrium point and the absolute values of the results are reported. The behaviour of PDM as a function of bond distance $R$, observed with CCSD(T)/QZ method, for both the molecular ions is shown in Figure~\ref{fig2}. Our results calculated at the equilibrium point are compared directly with those available in the literature and with those extracted from the dipole moment curves in the case of non-availability of these values. As the calculated values of PDM are fairly large, these ions might be useful for the study of long range dipole-dipole interactions.\\
%
\begin{figure}[ht]
\includegraphics[clip,width=\columnwidth]{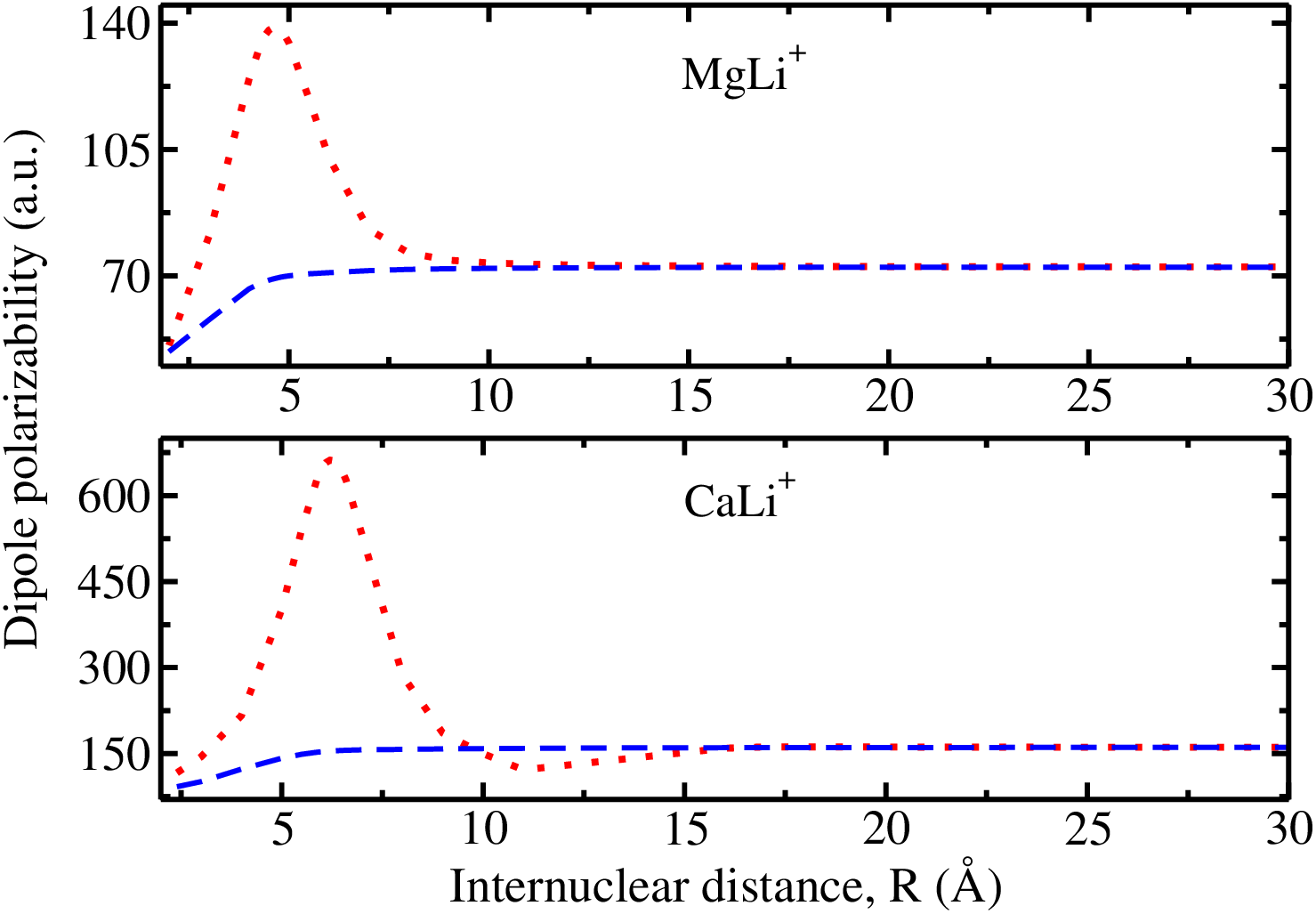}
\caption{\label{fig4} (colour online) Parallel (red dotted line) and perpendicular (blue dashed line) components of dipole polarizability  
using CCSD(T)/QZ method.}
\end{figure}
%
\begin{figure}[ht]
\includegraphics[clip,width=\columnwidth]{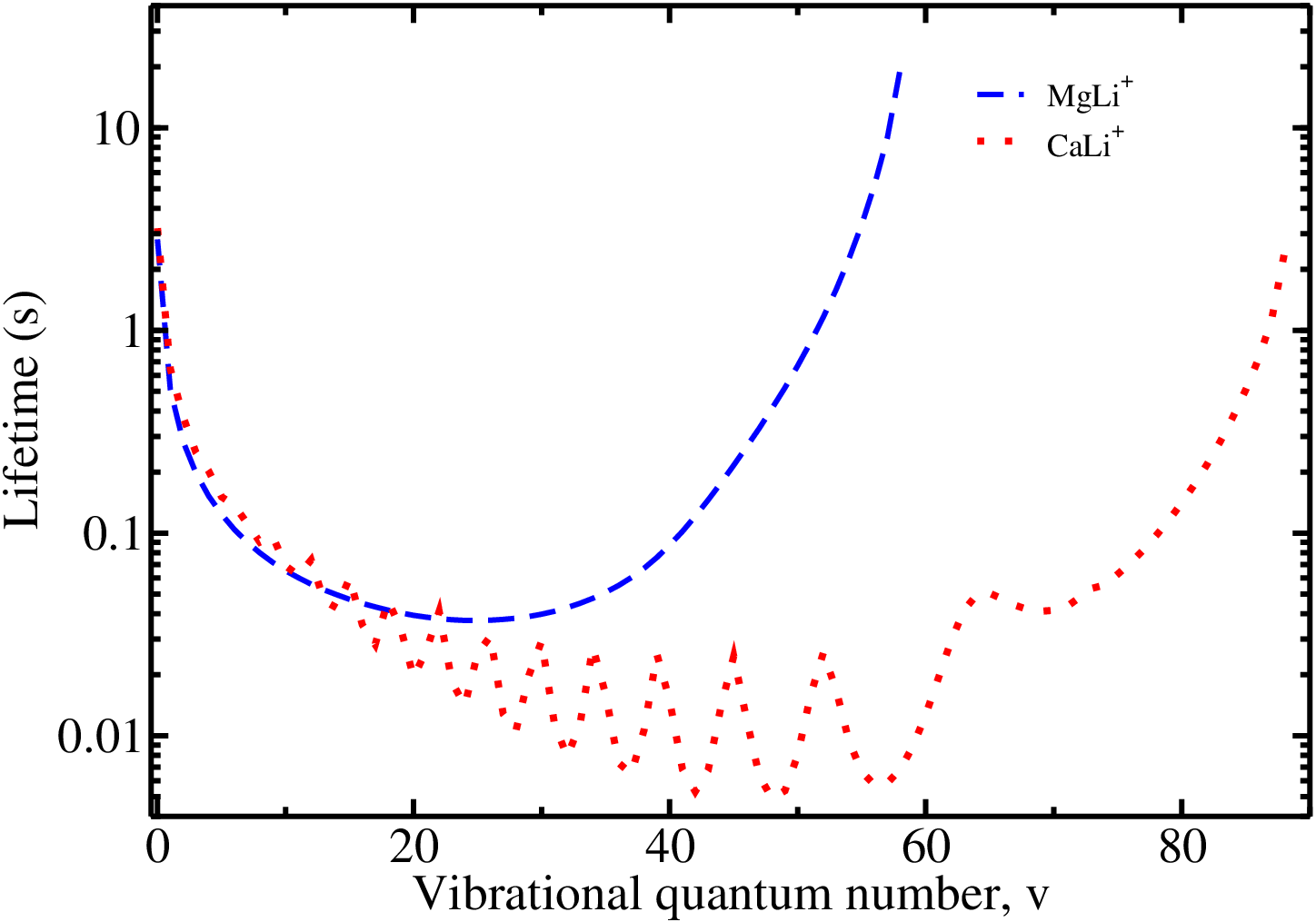}
\caption{\label{fig5} (colour online) Lifetimes of the vibrational states (with $J\,=\,0$) of the electronic ground state at T = $300$K.}
\end{figure}
%
%
\begin{table*}[ht]
\begin{ruledtabular}
\begin{center}
\caption{\label{table-III} The dipole- and traceless quadrupole moments together with the components of static dipole polarizabilities at the molecular equilibrium point and the latter at super molecular limit for the ground state of MgLi$^+$ and CaLi$^+$. The results are quoted in atomic units.}
\begin{tabular}{c c c c c c c c c}
Molecule & Method/basis & $\mu_z$  & $\Theta_{zz}$  & $\alpha$$_{\parallel }$  & $\alpha$$_{\bot }$  & $\bar{\alpha}$  & $\gamma$ & $\alpha$$_{100}$ \\
  \hline 
 MgLi$^+$ & SCF/TZ &  2.086 & 20.045 & 84.464 & 61.236 & 68.979 & 23.228 & 81.386\\
           & MP2/TZ &  2.112 & 19.308 & 79.383 & 57.217 & 64.606 & 22.165 & 73.387 \\
           & CCSD/TZ & 2.139 & 19.165 & 78.408 & 57.216 & 64.280 & 21.192 & 72.273\\
           & CCSD(T)/TZ &  2.146 & 19.122 & 78.145 & 56.975 & 64.032 & 21.170 & 71.823  \\
           \hline
           & SCF/QZ &  2.085 & 20.041 & 84.485 & 61.339 & 69.054 & 23.146 &  81.544\\
           & MP2/QZ &  2.103 & 19.214 & 78.642 & 57.128 & 64.299 & 21.514 & 73.652\\
           & CCSD/QZ &  2.129 & 19.075 & 77.885 & 57.151 & 64.062 & 20.734 & 72.938\\
           & \textbf{CCSD(T)/QZ} &   \textbf{2.135} &  \textbf{19.017} &  \textbf{77.533} &  \textbf{56.863} &  \textbf{63.753} &  \textbf{20.669} &  \textbf{72.462}\\
           & \textbf{Error bar} & $\pm$\textbf{0.017} &  $\pm$\textbf{0.163} & \textbf{$-$} &  \textbf{$-$} &  $\pm$\textbf{0.588} &  $\pm$\textbf{0.566} &  $\pm$\textbf{1.115}\\
           \hline
                      & CCSDT/cc-pCVQZ~\cite{Fedorov} &  2.140 &$-$&$-$&$-$& $-$ & $-$& $-$\\
                      & MRCI/cc-pCVQZ~\cite{Fedorov} &  2.140 &$-$&$-$&$-$& $-$ & $-$& $-$\\
           & MRCI/AV5Z+Q+3DK~\cite{Gao} & 2.126 & $-$ & $-$ & $-$ & $-$ &$-$ & $-$\\
           &  FCI/Gaussian~\cite{ElOualhazi} &  3.075 &$-$&$-$&$-$& $-$ & $-$& $-$\\
           \hline \hline
           
  CaLi$^+$ & SCF/TZ &  1.441 & 24.140 & 181.719 & 117.212 & 138.714 & 64.507 & 185.046\\
           & MP2/TZ &  1.827 & 23.401 & 165.102 & 102.065 & 123.077 & 63.037 & 144.279 \\
           & CCSD/TZ &  1.695  & 23.344 & 163.684 & 108.221 & 126.708 & 55.463 & 163.853  \\
           & CCSD(T)/TZ&  1.751 & 23.328 & 161.008 & 107.309 & 125.209 & 53.700 & 161.862 \\
           \hline
           &  SCF/QZ &  1.439 & 24.121 & 181.668 & 117.329 & 138.776 & 64.339 & 185.461\\
           & MP2/QZ &  1.814 & 23.182 & 162.911 & 101.711 & 122.111 & 61.199 & 143.741\\
           & CCSD/QZ &  1.685 & 23.157 & 162.699 & 107.947 & 126.198 & 54.752 & 163.386 \\
           &  \textbf{CCSD(T)/QZ} &   \textbf{1.748} &  \textbf{23.120} &  \textbf{159.409} &  \textbf{106.738} &  \textbf{124.295} &  \textbf{52.671} &  \textbf{160.812}\\
           & \textbf{Error bar} & $\pm$\textbf{0.066} & $\pm$\textbf{0.245} & \textbf{$-$} & \textbf{$-$} & $\pm$\textbf{2.817} & $\pm$\textbf{3.11} & $\pm$\textbf{3.624}\\
           \hline
           & FCI/Gaussian \cite{Habli}&  2.713 &$-$ & $-$ & $-$ & $-$ & $-$ & $-$\\
\end{tabular}
\begin{flushleft}
\end{flushleft}
\end{center}
\end{ruledtabular}
\end{table*}
The $R$\,-\,variation of quadrupole moment and components of dipole polarizability for both the molecular cations is shown in Figure~\ref{fig3} and Figure~\ref{fig4}, respectively. The behaviour of quadrupole moment and dipole polarizability curve is similar to that observed for BeLi$^+$ ion in Ref.~\cite{BeLi+}. Our recommended values of the results obtained at the highest level of correlation, i.e., CCSD(T) and with the largest basis sets, i.e., QZ, are highlighted, together with the error bars, in bold fonts at the bottom of Table~\ref{table-III} and they could serve as benchmarks for other calculations in future and also they may be useful for experimentalists who would consider working on these molecular systems in future. The results of quadrupole moments and electric dipole polarizabilities for these molecular ions are not available in the literature, to the best of our knowledge, to compare with.\\
We have also investigated the effect of adding diffuse functions, on the results of molecular properties by considering singly augmented QZ basis sets together with the CCSD(T) method. The corresponding results for MgLi$^+$ (CaLi$^+$) are: $\mu_e = 2.135 (1.749)$ a.u., $\Theta_{zz} = 19.015 (23.134)$ a.u., $\bar\alpha = 63.795 (124.404)$ a.u., $\alpha_{\bot} = 20.581 (52.550)$ a.u., and $\alpha_{100} = 73.099 (160.882)$ a.u. It is thus observed that the contribution of diffuse functions to the dipole-, quadrupole- moments and polarizabilities is well within the error bars quoted on these properties in Table~\ref{table-III}.\\
In order to determine the molecular products to which the ground state of MgLi$^+$ and CaLi$^+$ dissociate into, at asymptotic limits, we have calculated the  first ionization energies by taking the difference between the energies of neutral and singly charged molecules at $100$\,a.u. bond distance. The first ionization energy of both MgLi and CaLi molecules is the same, {\itshape viz.} $43437.6$ cm$^{-1}$,  and it matches well with the value of ionisation energy of Li atom ($43487.1$\,cm$^{-1}$), reported in NIST database~\cite{NIST} to within $0.11$\%. Hence, we realize that the ground state of these ions will dissociate to X(ns$^2$)\,+\,Li$^+$ with n\,=\,$3$ for Mg and $4$ for Ca. On solving the vibrational Schr\"odinger equation using PEC and PDM curve obtained at CCSD(T)/QZ level, we have obtained 59 and 89 vibrational states for MgLi$^+$, and CaLi$^+$, respectively. The energy separation between the last two vibrational levels in both cases is less than $0.3$ cm$^{-1}$. The relative vibrational energies, rotational constants and TDMs between different vibrational states for the electronic ground state of MgLi$^+$ and CaLi$^+$ are reported in the Supplementary Tables S1 and S2, respectively.   
The calculated lifetime of the electronic ground state with $v=0$ and $J=0$ is $2.81\,\mathrm s$ for MgLi$^+$ and $3.19\,\mathrm s$ for CaLi$^+$. The variation of the lifetimes of vibrational states against vibrational quantum number, $v$ is shown in Figure~\ref{fig5}. \\
In the following subsections, we will discuss the spectroscopic constants and molecular properties for the electronic state of individual ionic systems considered in this work in detail and compare them with the available calculations. 
\begin{figure}[ht]
	\includegraphics[clip,width=\columnwidth]{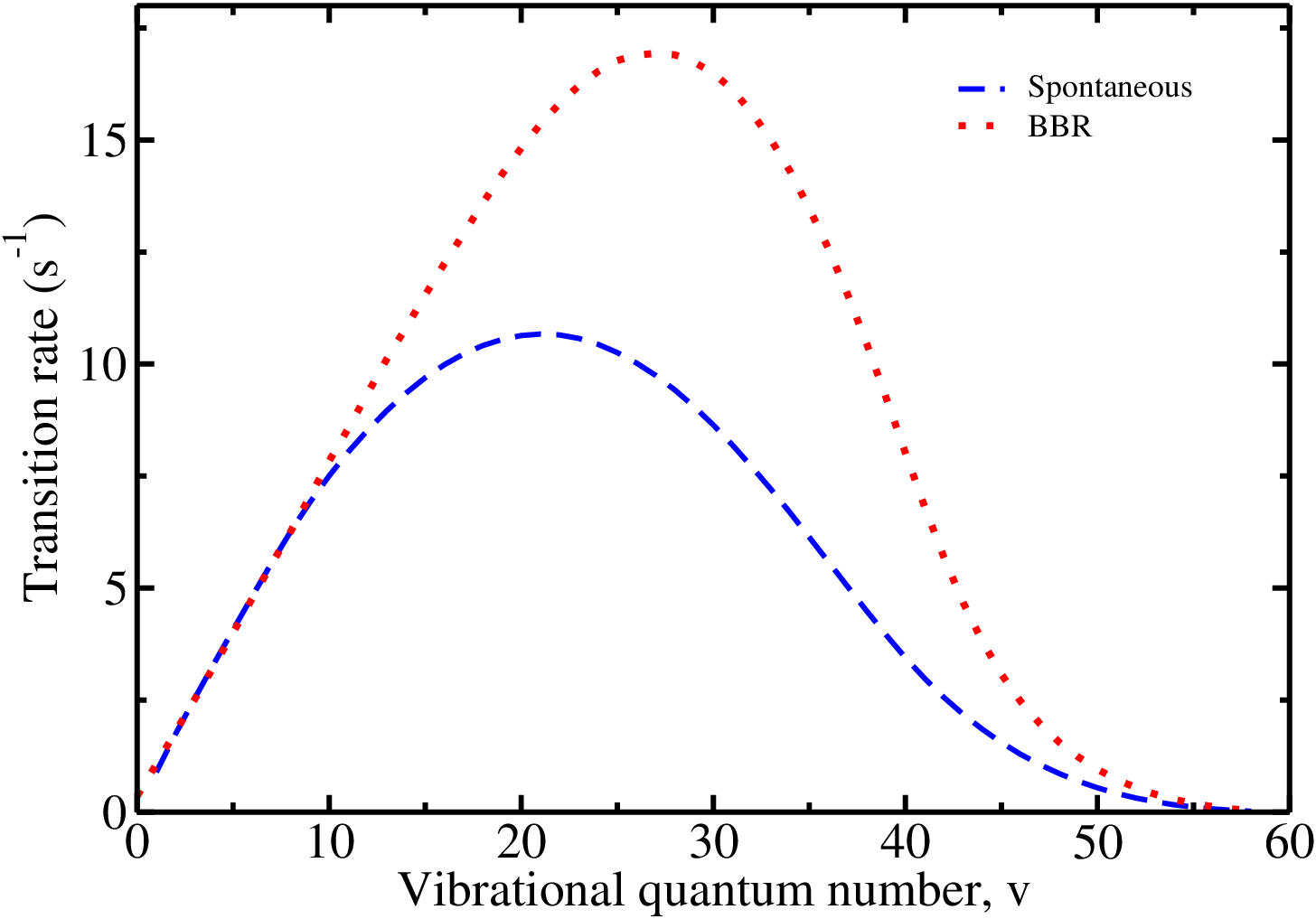}
	\caption{\label{fig6}(colour online) Spontaneous and BBR transition rates for the vibrational levels of MgLi$^+$.}
\end{figure}
%
\subsubsection{MgLi$^+$}
The results reported in Refs.~\cite{Pyykko, Boldyrev, Fantucci} for $R_e$, $D_e$ and $\omega_e$ at SCF and MP2 levels differ from ours, at the similar level of approximation, by a maximum of about $1.2$\%, $3$\%, and $1.8$\%, respectively, despite the fact that they have used minimal basis sets as against our QZ basis sets which are quite large in comparison. Our recommended value of $R_e$, {\itshape viz.} ($5.493\pm0.02$)\,a.u., as shown in Table \ref{table-I}, and the MRCI result quoted in Ref.~\cite{Gao} and also the FCI result of Ref.~\cite{ElOualhazi}, agree extremely well to within $1$\%. Similar is the situation for all other diatomic constants which have a maximum difference of $4$\% to ours, except the anharmonic frequency for which the FCI result is larger than ours by about 14\%. Our other results at CCSD(T)/QZ level differ from those reported in Ref.~\cite{Fedorov} at CCSDT and MRCI level with cc-pCVQZ basis set by less than 1\%. From the non-relativistic and relativistic calculations shown in Table~\ref{table-I} and Table~\ref{table-II}, respectively, we observe that the relativistic terms seem to have the effect of lowering the values of the spectroscopic constants that are reported. The value of $R_e$ reduces by about $0.05$\%; the dissociation energy $D_e$ reduces by about $0.24$\%, the harmonic frequency $\omega_e$ decreases by $0.6$\% and anharmonic constant $\omega_ex_e$ decreases by $8$\%. \\ 
Our value of the dipole moment, given in Table \ref{table-III}, of MgLi$^+$ at the equilibrium point in its ground state is ($2.135\pm0.017$)\,a.u., at the CCSD(T)/QZ level, and this compares well, to within $0.4$\% and $0.2$\%, with that of \cite{Gao} and \cite{Fedorov}, respectively. However, these results, ours included, differ significantly, by about $44$\%, from that given in Ref.~\cite{ElOualhazi}. 
The results of quadrupole moment and the static dipole polarizabilities for this ion are also reported in this work, most likely for the first time. The convergence trend in the correlation contributions may be seen in several molecular properties reported in Table \ref{table-III}. With reference to the SCF result, the contribution of the total electron correlations computed at the CCSD(T) level for $\Theta_{zz}$, $\bar{\alpha}$, and $\gamma$ is $5.1\%$, $7.7\%$ and $10.7\%$, respectively. Further, we observe that the contribution of leading order triples to the CCSD values of $\Theta_{zz}$ and $\gamma$ is $\sim0.3\%$, while for $\bar{\alpha}$ it is  $\sim0.5\%$. The value of polarizability at the super molecular limit in our calculation is equal to $\,72.462\,$\,a.u., with an error bar of $\pm\,1.115$,\,a.u. and it compares well with the sum of the atomic polarizabilities, $\alpha_{Mg}\,+\,\alpha_{Li^+}\,=\,71.37\,+\,0.191\,=\,71.561$\,a.u.~\cite{Miadokova}, at the similar level of approximation. \\
 In this work, we have obtained the maximum number of vibrational states to be $59$, whereas, Fedorov \emph {et al.}~\cite{Fedorov} have reported $54$. To analyze this further, we have considered the PEC upto  $R$ ($=20$ {\AA}), as is done in Ref.~\cite{Fedorov}, and  obtained exactly the same number of vibrational states, \textit {viz.} 54 despite the potential depths being different in both cases, the latter being $\sim 54\, $cm$^{-1}$ smaller. As it can also be seen in Ref.~\cite{Fedorov}, the maximum number of vibrational states is the same, though the difference between the dissociation energies calculated using CCSDT/cc-pCVQZ and MRCI/aug-cc-pCV5Z methods is 18 cm$^{-1}$. The relative energy difference between the adjacent vibrational levels in our case is similar to that reported in Ref.~\cite{Fedorov}, and the last difference being $\sim 2$ cm$^{-1}$ in both cases. However, by using the PEC upto $30$ {\AA}, the distance that we have considered as the dissociation limit, we have obtained $57$ bound vibrational states, that is, an addition of three vibrational states within the potential energy difference of 3.15 cm$^{-1}$,  between $R =20$ {\AA} and $R = 30$ {\AA}. The relative energy difference between the last two vibrational states is now  reduced to $0.91$ cm$^{-1}$, {\textit viz.} between the states  $v=56$ and $v=55$. On extending the PEC further from $30$ {\AA} to $52$ {\AA}, the latter being a  conventional dissociation limit, we have got two more vibrational states to make the total count to 59. Thus, the difference in the electronic energies of about $3.83$ cm$^{-1}$ between $R=20$ {\AA} and $R=52$ {\AA} increases the number of states from $54$ to $59$.\\ 
 The lifetime of the rovibronic ground state {\itshape viz.} $2.81\,\mathrm s$, reported in this work agrees well with the result, {\itshape viz.} $2.76\,\mathrm s$ reported in Ref.~\cite{Fedorov}. The lifetimes of the higher vibrational levels close to dissociation limit are observed to be longer than that of the rovibronic ground state. To analyze this further, we have shown the variation of spontaneous and BBR transition rates for the vibrational states of MgLi$^+$ against the vibrational quantum number in Figure~\ref{fig6}. The spontaneous and the BBR transition rates increase initially and they reach the peak at $v=21$ and at $v=27$, respectively. Thereafter, both the transition rates begin to decline. As the lifetime is the reciprocal of total transition rate, the lifetimes of the vibrational levels decrease upto $v\,=\,25$ and then they begin to increase. For $v=58$, the lifetime reaches the highest value of $19.3\,\mathrm s$, which is about seven times larger than the lifetime of $v\,=\,0$ state. For both BBR-induced and spontaneous transition rates, we have observed that the transitions with $ \Delta v \,=\, \vert 1 \vert$ are dominant for the lower vibrational levels. However, for intermediate and higher vibrational levels, the fundamental as well as the first few overtones contribute appreciably. The long lifetimes for highly excited vibrational states have also been predicted by Fedorov et al.~\cite{Fedorov} for alkali-alkaline-earth cations and by Zemke~\emph{et al.} \cite{Zemke} for neutral KRb molecule.\\
 \begin{figure}[ht]
 	\includegraphics[clip,width=\columnwidth]{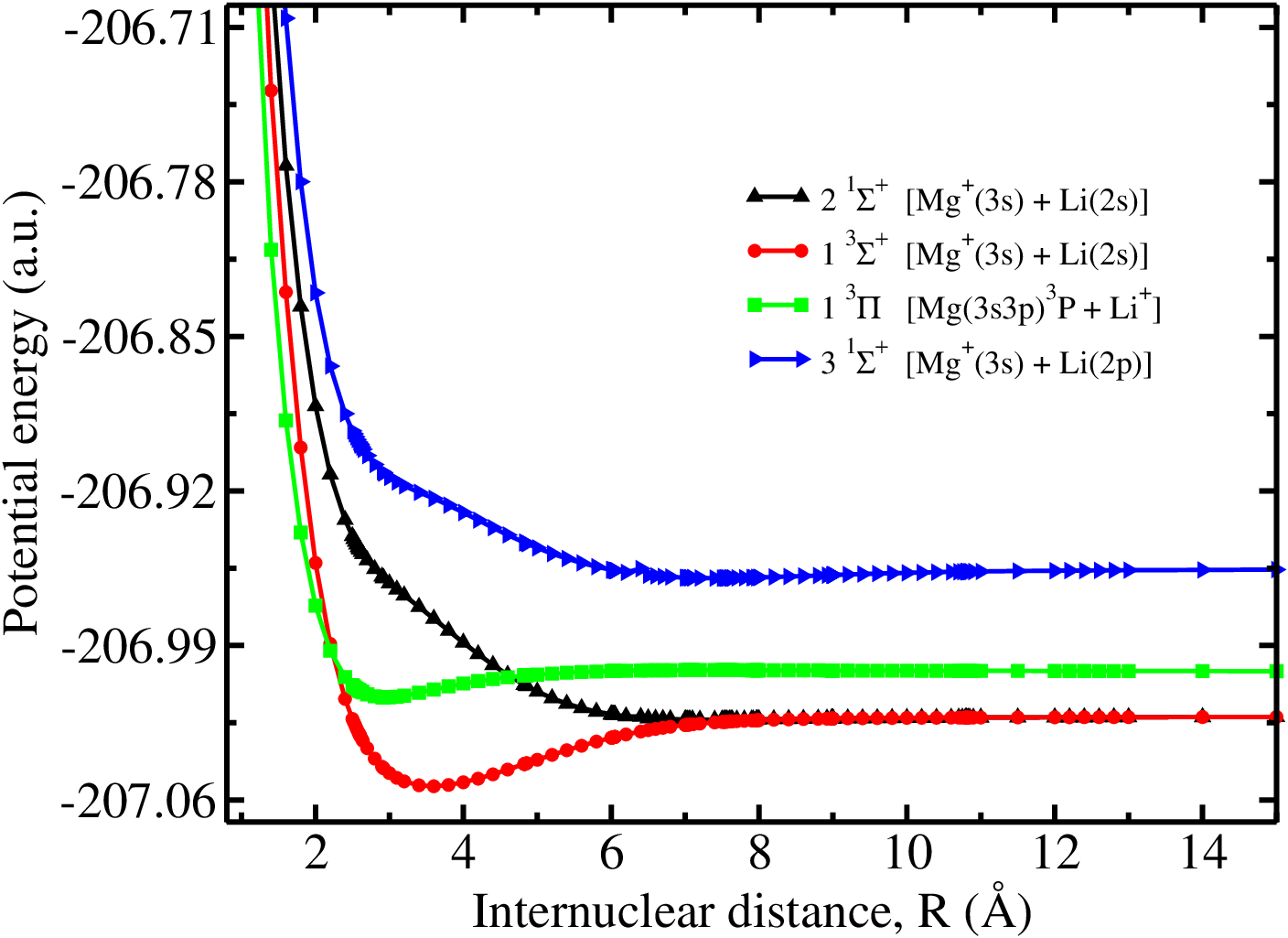}
 	\caption{\label{fig7}(colour online) Potential energy curves for the electronic excited states of MgLi$^+$ using EOM-CCSD/QZ method.}
 \end{figure}
\subsubsection{CaLi$^+$}
The results of $R_e$, $D_e$ and $\omega_e$ for the ground state of CaLi$^+$ molecular ion, computed at the equilibrium bond length using QCISD(T) method, reported by Russon~\emph{et al.}~\cite{Russon} differ from our CCSD(T) results, shown in Table~\ref{table-I}, by about -$1.7$\%, $0.8$\% and $2.5$\%, respectively. However, their QCISD(T,full) result of $D_e$ differs from our result by $4.1$\%. On the other hand, our results for $R_e$, $D_e$ and $\omega_e$ agree very well with those of Habli \emph{et al.}~\cite{Habli} with a maximum disagreement being $1.26\%$. The value of $D_e = 8952.8$\, cm$^{-1}$ reported in~\cite{Kimura} is quite small compared to all other published results and it is about $11.3$\% smaller than our result. However, our value of the rotational constant, $B_e = (0.265\pm0.002)$\,cm$^{-1}$ agrees nicely with that of~\cite{Kimura} where it is quoted as $0.263$\,cm$^{-1}$. The anharmonic frequencies are not available in the literature for comparison. From Table~\ref{table-II}, we observe that the relativistic effects, at the CCSD(T)/QZ level, on $R_e$ is a mere -$0.08$\%, on $D_e$ and on $\omega_e$ it is about - $0.8$\%, and on $\omega_ex_e$ it is about -$15$\%. \\ 
The only available result of the dipole moment for CaLi$^+$ in the literature is by Habli~\emph{et al.}~\cite{Habli} which reports its value to be $\sim 2.713$~a.u. at the equilibrium bond length. This is larger than our CCSD(T) result, ($1.748\pm0.066$)\,a.u. There is no calculation available in the literature, known to our knowledge, which reports the results of quadrupole moment and static dipole polarizabilities for this ion and hence, our results reported in this work are most likely the first. The sum of individual atomic polarizabilities, $\alpha_{Ca}\,+\,\alpha_{Li^+}\,=\,155.9\,+\,0.191\,=\,156.091$\,a.u. reported in Ref.~\cite{Miadokova} is very close to our result of polarizability ($160.812\,\pm\,3.624$\,a.u.) calculated at the super-molecular limit. The correlation contributions due to CCSD and CCSD(T) to SCF, are positive for $\mu_z$, while they are negative for the other calculated molecular properties, similar to the trends observed in the case of MgLi$^+$. \\ 
The computed lifetime of the rovibronic ground state of CaLi$^+$, at room temperature, is found to be $3.19\,\mathrm s$. There is no other data available in the literature for comparison. From Figure~\ref{fig5}, we observe that the nature of the curve showing the lifetime of vibrational states of CaLi$^+$ against $v$ is similar to the case of MgLi$^+$. The lifetime for the highest vibrationally excited state is found to be $2.38\,\mathrm s$, which is close to the value of the lifetime of the  vibrational ground state. The behaviour of the BBR transition rate with $v$ is analogous to that observed for MgLi$^+$, however, the spontaneous rate curve shows several minor oscillations between $v=9$ and $v=70$ and the same gets reflected in the lifetime curve as well.\\ 
%
\begin{table*}[ht]
\begin{ruledtabular}
\begin{center}
\caption{\label{table-IV} The calculated spectroscopic constants for some lower excited states of MgLi$^+$ using EOM-CCSD/QZ method, compared with the available results in the literature.}
\begin{tabular}{l l l l l l l l l}
State & $R_e$~({\emph {a.u.})} & $D_e$ (cm$^{-1}$) & $T_e$ (cm$^{-1}$) & $\omega$$_e$(cm$^{-1}$) & $\omega$$_ex_e$(cm$^{-1}$) & $B_e$ (cm$^{-1}$) & $\alpha_e$ (cm$^{-1}$) & Ref.\\
 \hline
  2$^1\Sigma^+$ & 13.795	&	283.1	&	24618.03	&	39.955	&	1.97	&	0.0595	&	0.0031	 & This work\\
                & 10.73 & 1248 & 23647 & 79.59 & 1.12 & 0.096860 & $-$ &  \cite{ElOualhazi}\\
                  \hline																			
1$^3\Sigma^+$  & 6.752	& 6908.1	& 17992.74	& 179.490	& 0.45	& 0.2347 & 0.0007	 & This work\\                 
               & 6.712 & 7668.2 & 16339.1 & 188.8 & 0.90 & 0.2462 &$-$ &  \cite{Gao}$^a$\\
               & 6.701 & 7679.4 & 16441.5 & 187.8 & 0.84 & 0.2470 &$-$ &  \cite{Gao}$^b$\\
               & 6.705 & 7678.3 & 16292.2 & 189.4 & 0.92 & 0.2468 & $-$ &  \cite{Gao}$^c$\\
               & 6.703 & 7668.5 & 16355.6 & 189.3 & 0.92 & 0.2468 & $-$ &  \cite{Gao}$^d$\\
               & 6.64 & 7983 & 16912 & 189.96 & 1.43 & 0.252539 & $-$ &  \cite{ElOualhazi}  \\
                  \hline
  1$^3\Pi$ &  5.584	& 2578.6	&	26824.62	& 215.442	& 4.38	& 0.3567	& 0.0080	 & This work\\      
                    & 5.670 & 2782.6 & 24977.6 & 208.6 & 3.46 & 0.3452 &$-$ &  \cite{Gao}$^a$\\
          & 5.650 & 2822.9 & 25070.0 & 212.0 & 3.61 & 0.3475 &$-$ &  \cite{Gao}$^b$\\
          & 5.671 & 2742.3 & 24967.7 & 208.1 & 3.46 & 0.3449 &$-$ &  \cite{Gao}$^c$\\
          & 5.670 & 2726.2 & 25053.1 & 207.8 & 3.47 & 0.3451 &$-$ &  \cite{Gao}$^d$\\
          & 5.60 & 2561 & 26008 & 206.32 & 3.51 & 0.356099& $-$ &  \cite{ElOualhazi}\\
                              \hline
  3$^1\Sigma^+$ & 14.000	& 938.9 &	38691.4	&	61.7308	&	1.39	&	0.0573	&	0.0011 & This work\\
                & 12.58 & 2548 & 37252 & 70.38 & 0.48 & 0.070509 & $-$ &  \cite{ElOualhazi}\\       
\end{tabular}
\begin{flushleft}
$^a$MRCI/AV5Z\,+\,Q, $^b$MRCI/AV5Z\,+\,Q\,+\,DK, $^c$MRCI/AVQZ\,+\,Q, $^d$ MRCI/AVQZ\,+\,Q\,+\,DK, 
\end{flushleft}
\end{center}
\end{ruledtabular}
\end{table*}
%
\begin{table*}[ht]
\begin{ruledtabular}
\begin{center}
\caption{\label{table-V} Energies of a few low-lying electronic states of MgLi$^+$ at the dissociative limit.}
\begin{tabular}{l l l l l l l l l}
     & & \multicolumn{2}{c}{E (cm$^{-1}$)} &\\
                   \cline{3-4}  \\
  Molecular state & Asymptotic molecular state & This work & NIST~\cite{NIST} & \% error\\
    \hline 
    X$^1\Sigma^+$ & Mg\,(3s$^2$)\,+\,Li$^+$& 0.0 & 0.0 & $-$\\
    2$^1\Sigma^+$ & Mg$^+$\,(3s)\,+\,Li(2s) & 18198.24 & 18183.94 & 0.08\\
    1$^3\Sigma^+$ & Mg$^+$\,(3s)\,+\,Li(2s) & 18198.25 & 18183.94 & 0.08\\
    1$^3\Pi$ & Mg\,(3s\,3p)\,$^3$P\,+\,Li$^+$& 22698.04 & 21877.23 & 3.75\\
    3$^1\Sigma^+$ & Mg$^+$\,(3s)\,+\,Li(2p) & 32926.93 & 33087.77 & 0.49 \\
\end{tabular}
\begin{flushleft}
\end{flushleft}
\end{center}
\end{ruledtabular}
\end{table*}
%
\begin{figure}[ht]
\includegraphics[clip,width=\columnwidth]{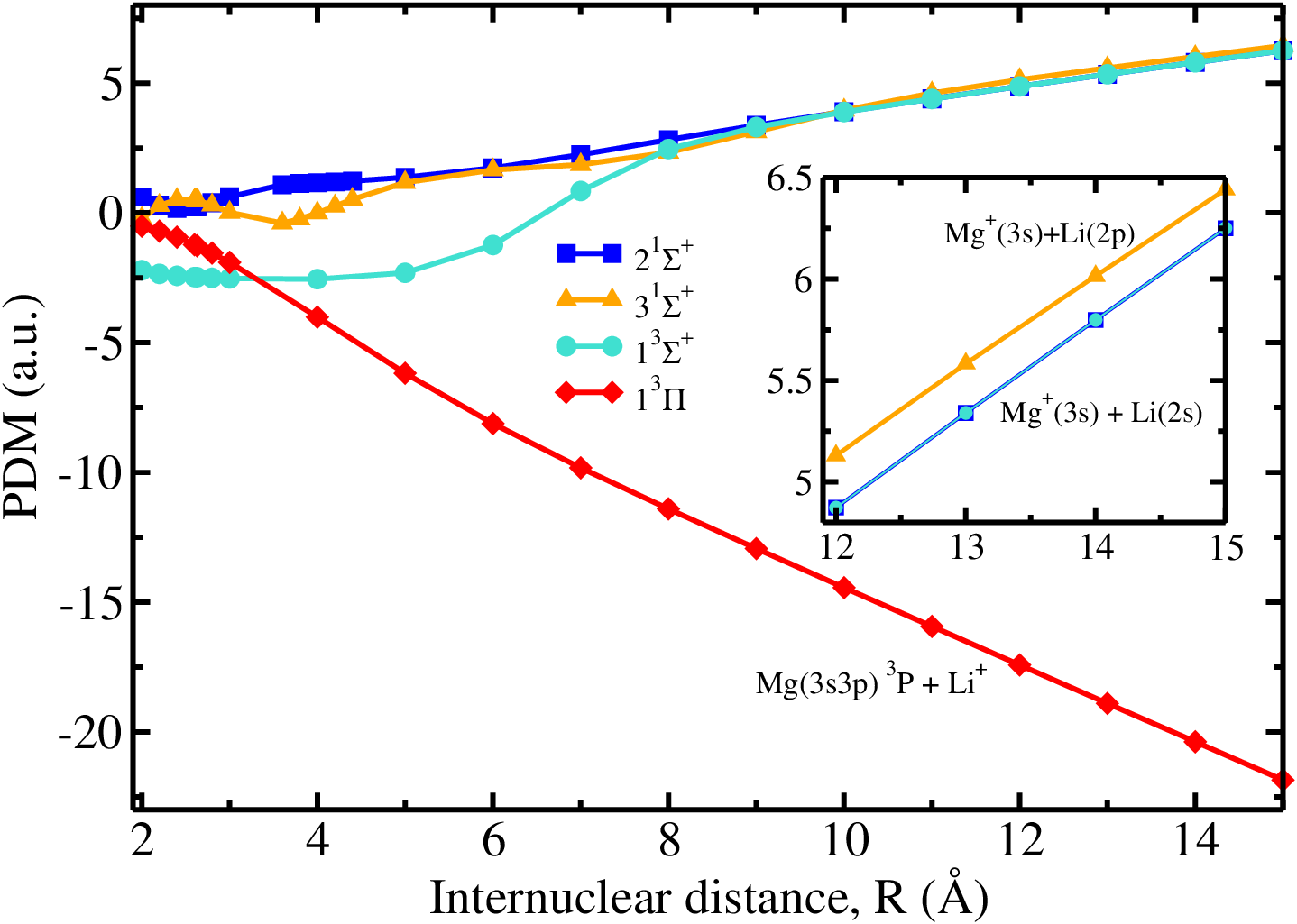}
\caption{\label{fig8}(colour online) Permanent dipole moment curves for the excited states of MgLi$^+$ using EOM-CCSD/QZ method.}
\end{figure}
%
\begin{figure}[ht]
\includegraphics[clip,width=\columnwidth]{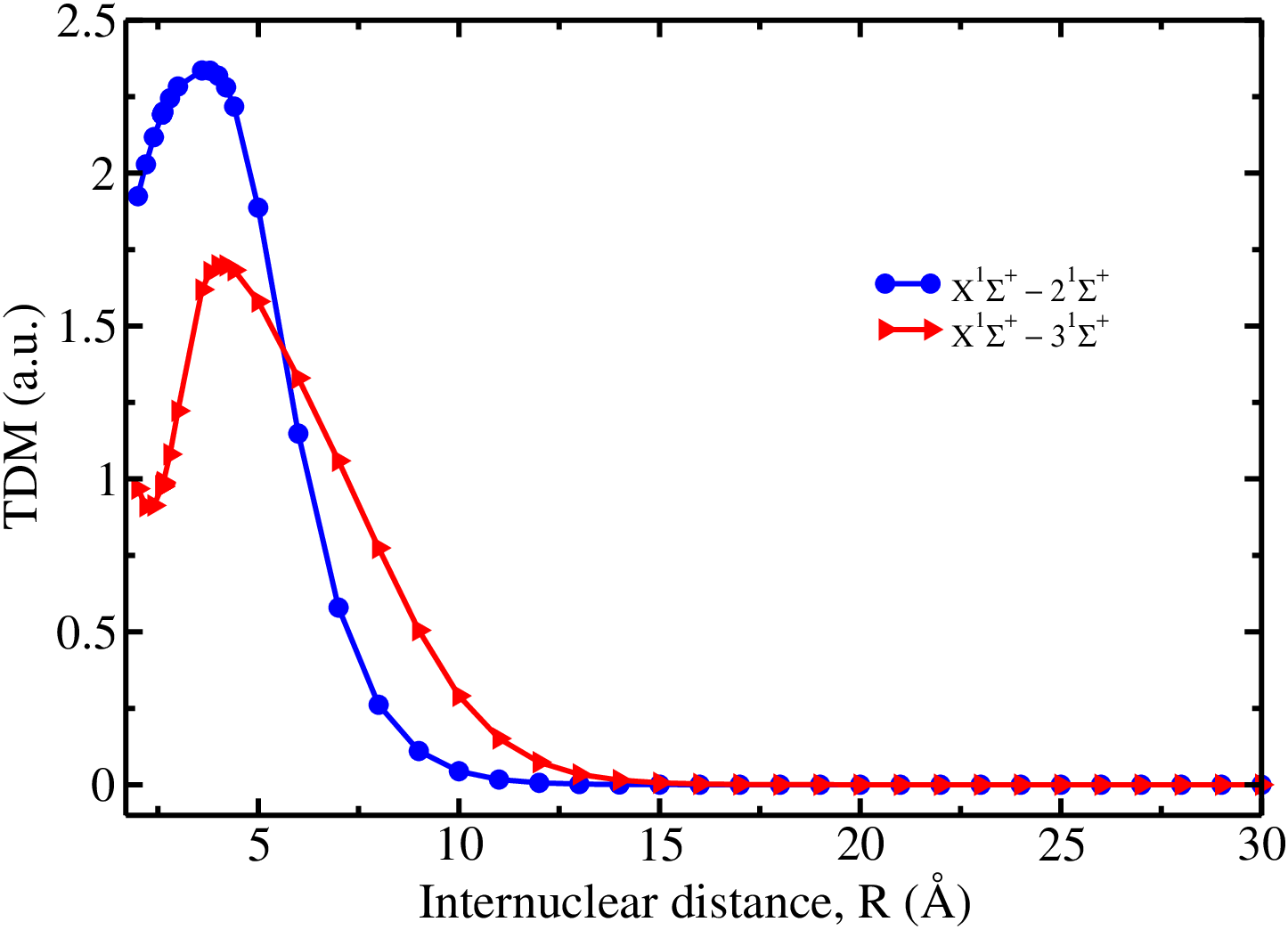}
\caption{\label{fig9}(colour online) TDM curves for the transitions from the electronic ground- to the singlet-excited states of MgLi$^+$ using EOM-CCSD/QZ method.}
\end{figure}
\subsection{Excited state spectroscopic constants and molecular properties}
\subsubsection{MgLi$^+$}
\begin{figure}[ht]
\includegraphics[clip,width=\columnwidth]{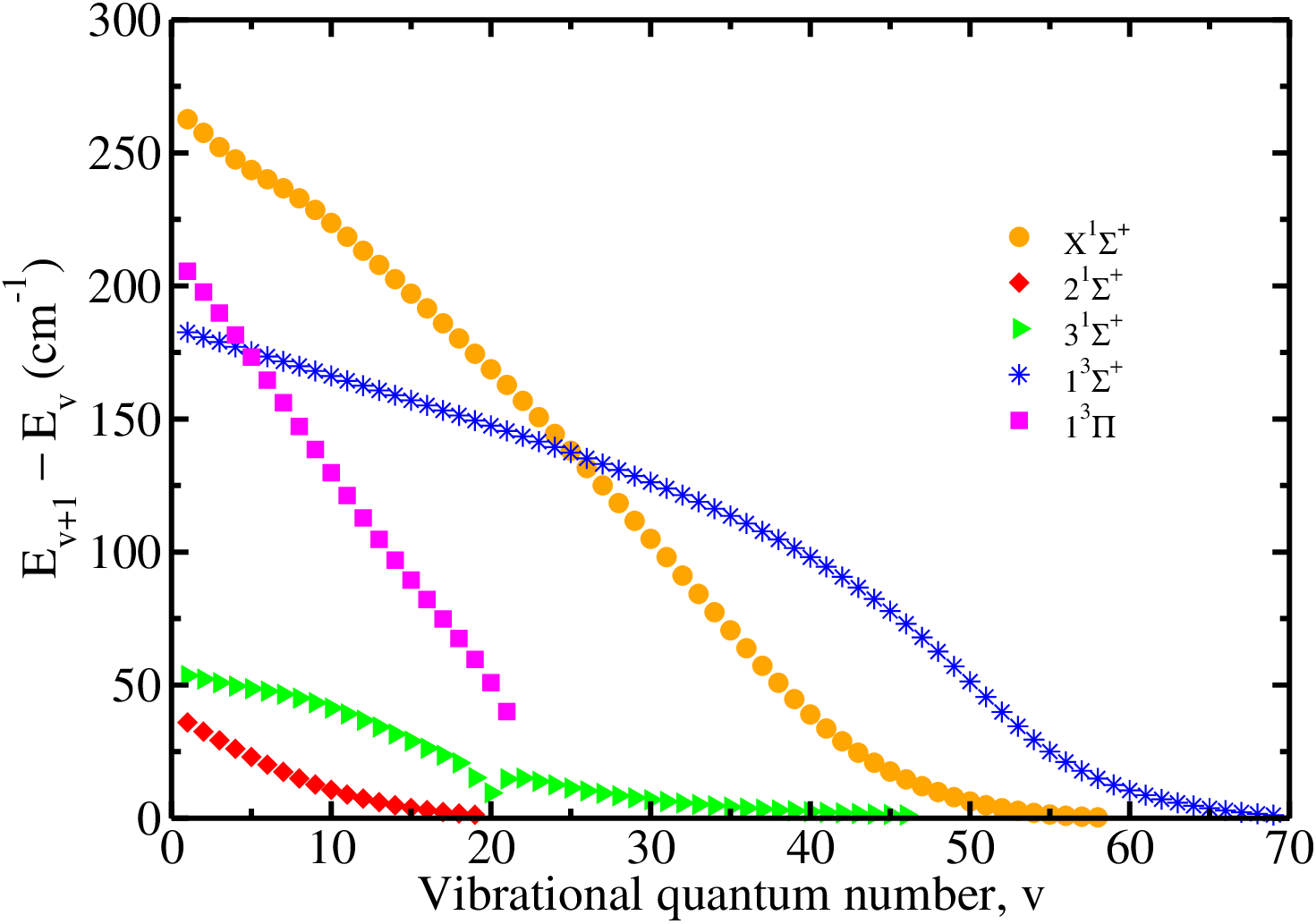}
\caption{\label{fig10}(colour online) Energy spacing between the adjacent vibrational levels of the  electronic ground- and of the excited states of MgLi$^+$.}
\end{figure}

The PECs and PDM curves for the excited states of MgLi$^+$: 2\,-\,3$^1\Sigma^+$, 1$^3\Sigma^+$, and 1$^3\Pi$, at EOM-CCSD/QZ level of the theory are shown in Figure~\ref{fig7} and Figure~\ref{fig8}, respectively. The spectroscopic constants extracted from these PECs are tabulated in Table~\ref{table-IV} and compared with the published results available in the literature. The spectroscopic constants: $R_e$, $\omega_e$, and $B_e$, for the triplet excited states reported in this work are reasonably in good agreement with the other calculations ~\cite{Gao, ElOualhazi}. The value of $\omega_e x_e$, on the other hand, shows a noticeable difference to the published results. The spectroscopic constants for the singlet n$^1\Sigma^+$ excited states show large disagreement with those of Ref.~\cite{ElOualhazi} mainly because our value of $R_e$ is largely different from that considered in their work which renders the direct comparison of the two results meaningless. The dissociation energy $D_e$, and the electronic transition energy $T_e$ for the triplet excited states reported in our work differ from those of published results for the following reasons. As seen from the results of Ref.~\cite{Gao} cited in Table~\ref{table-I} and Table~\ref{table-IV}, the magnitude of these energies are sensitive to the size of the basis set and also to the relativistic effects. Further, in Refs.~\cite{Gao, ElOualhazi} they have performed frozen core calculations as against our all-electron correlation calculations. The other reason could be the choice of the many-body method used in our work which is different from the other two.\\ 
We have observed that the two of the excited molecular states: 2$^1\Sigma^+$ and 1$^3\Sigma^+$ dissociate into [Mg$^+$\,(3s)\,+\,Li\,(2s)] states, 1$^3\Pi$ into [Mg(3s\,3p)\,+\,Li$^+$] states, and 3$^1\Sigma^+$ into [Mg$^+$\,(3s)\,+\,Li\,(2p)] states. The Refs.~\cite{Gao, ElOualhazi} also report the same observations on the dissociative nature of these molecular states. These asymptotic molecular states and the corresponding electronic excitation energies with respect to the molecular ground state is given in Table~\ref{table-V} and compared with the sum of atomic/ionic energies taken from the NIST~\cite{NIST} database. The maximum difference between our results and those of ~\cite{NIST} is $3.8$\%. On correlating Figure~\ref{fig7} and Figure~\ref{fig8}, we observe that the molecular states having positive values of dipole moment at large distances, dissociate into Mg$^+$\,(3s)\,+\,Li\,(2s, {\itshape or} 2p), whereas the states with negative value of dipole moment at large distances, dissociate into Mg\,(3s$^2$, {\itshape or} 3s3p)\,+\,Li$^+$.\\
The TDM curves for the transitions from the electronic ground- to singlet-excited states, are shown in Figure~\ref{fig9}. The transitions,
X$^1\Sigma^+$ $\rightarrow$ 2$^1\Sigma^+$ and X$^1\Sigma^+$ $\rightarrow$ 3$^1\Sigma^+$ have a maximum value of TDM at a distance of $3.6$ {\AA} and $4$ {\AA}, respectively. Further, the number of vibrational states obtained within 2$^1\Sigma^+$,  1$^3\Sigma^+$, 1$^3\Pi$, and 3$^1\Sigma^+$ electronic states are $20$, $70$, $22$, and $47$, respectively. The relative energy separation between the last two vibrational states for 2$^1\Sigma^+$, 3$^1\Sigma^+$, and 1$^3\Sigma^+$ is less than $1.2$ cm$^{-1}$, whereas, it is about $40$ cm$^{-1}$ for 1$^3\Pi$. We have plotted the energy spacing between adjacent vibrational states against $v$ for the ground as well as for the excited electronic states in Figure~\ref{fig10}. These results together with the vibrationally coupled rotational constants for the excited electronic states are provided in the Supplementary Table S3.
%
%
\begin{table*}[ht]
\begin{ruledtabular}
\begin{center}
\caption{\label{table-VI} The calculated spectroscopic constants for the low-lying excited states of CaLi$^+$, using EOM-CCSD/QZ method.}
\begin{tabular}{l l l l l l l l l}
 State & $R_e$~({\emph {a.u.})} & $D_e$ (cm$^{-1}$) & $T_e$ (cm$^{-1}$) & $\omega$$_e$(cm$^{-1}$) & $\omega$$_ex_e$(cm$^{-1}$) & $B_e$ (cm$^{-1}$) & $\alpha_e$ (cm$^{-1}$)  & Ref.\\
    \hline
   2$^1\Sigma^+$ & 22.41 & 23.39& 16500.54& 6.474&0.63& 0.0201& 0.0020& This work\\
                  & 13.68 & 412.37 & $-$ & 41 & $- $ & $- $   & $- $   &  \cite{Habli}\\
    \hline
    1$^3\Sigma^+$ & 7.609 & 3596.4 & 12925.95 & 141.873 & 0.22 & 0.1663 & 0.0003  & This work\\
                  & 7.18  & 5791.92& $-$    & 136     & $- $ & $- $   & $- $    & \cite{Habli}\\
    \hline
    1$^3\Pi$ & 5.879 & 7621.13 & 19933.56 & 255.875 & 1.57 & 0.2904 & 0.0021 & This work\\
             & 5.71 & 8982.19 & $-$ & 251 & $- $ & $- $   & $- $  & \cite{Habli}\\
    \hline
     1$^1\Pi$ & 6.472 & 1371.90 & 34547.72 & 214.752&1.66& 0.2397&0.0014& This work\\
              & 6.69 & 6643.47 & $-$ & 193 & $- $ & $- $   & $- $   &  \cite{Habli}\\            
\end{tabular}
\begin{flushleft}
\end{flushleft}
\end{center}
\end{ruledtabular}
\end{table*}
%
\begin{figure}[ht]
\includegraphics[clip,width=\columnwidth]{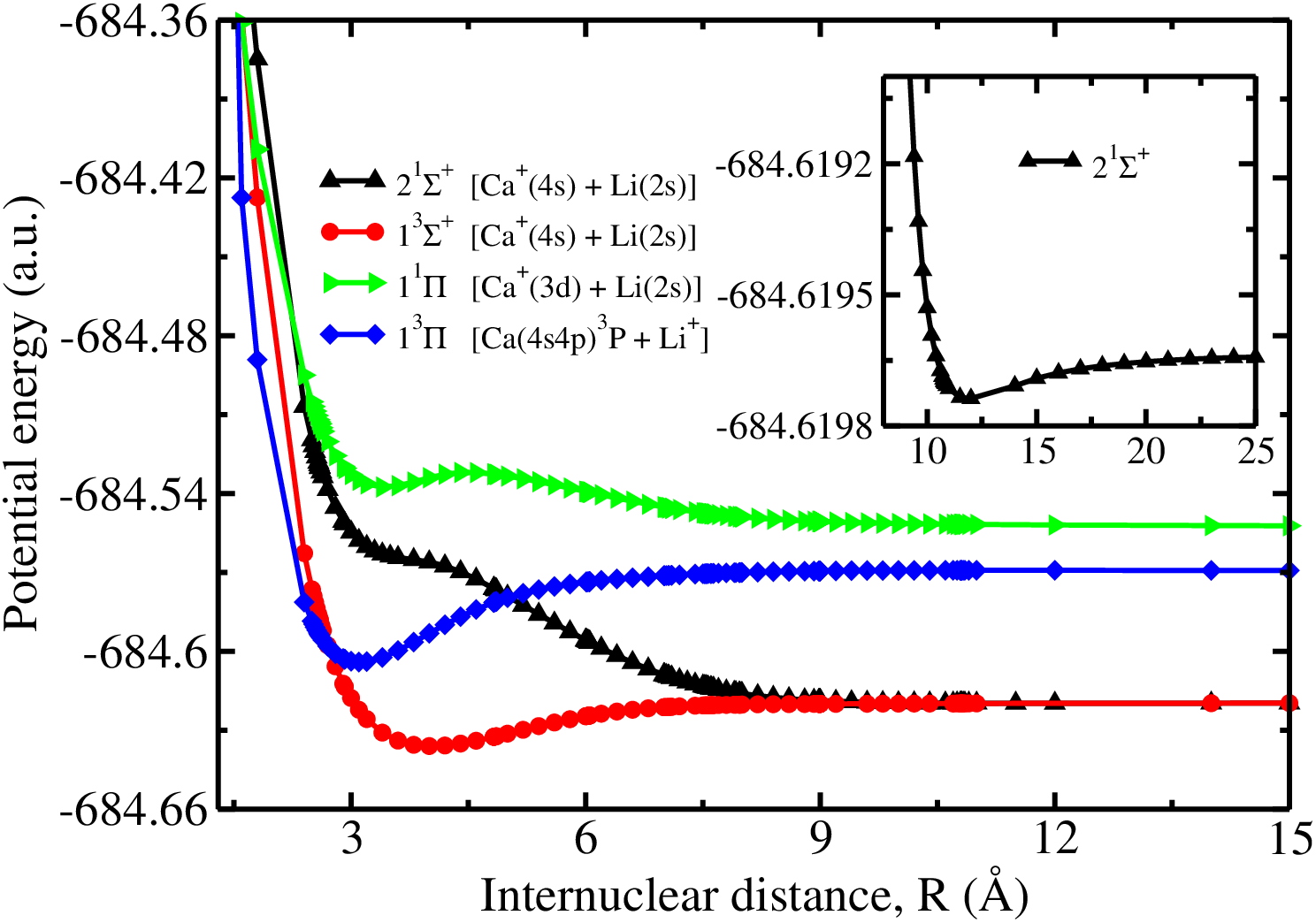}
\caption{\label{fig11}(colour online) Potential energy curves for the electronic excited states of CaLi$^+$ using EOM-CCSD/QZ method.}
\end{figure}
%
\begin{table*}[ht]
\begin{ruledtabular}
\begin{center}
\caption{\label{table-VII} Energies of a few low-lying electronic states of CaLi$^+$ at the dissociative limit.}
\begin{tabular}{l l l l l l l l l}
     & & \multicolumn{2}{c}{E (cm$^{-1}$)} &\\
                   \cline{3-4}  \\
  Molecular state & Asymptotic molecular state & This work & NIST~\cite{NIST} & \% error\\
    \hline 
    X$^1\Sigma^+$ & Ca\,(4s$^2$)\,+\,Li$^+$& 0.0 & 0.0 & \\
    2$^1\Sigma^+$ & Ca$^+$\,(4s)\,+\,Li(2s) & 6438.22 & 5818.82 & 10.6\\
    1$^3\Sigma^+$ & Ca$^+$\,(4s)\,+\,Li(2s) & 6438.22 & 5818.82 & 10.6\\
    1$^3\Pi$ & Ca\,(4s\,4p)\,$^3$P\,+\,Li$^+$& 17466.91 & 15227.97 & 15\\
    1$^1\Pi$ &  Ca$^+$\,(3d)\,+\,Li(2s) &  21170.74 & 19499.35 &  8.6\\
\end{tabular}
\begin{flushleft}
\end{flushleft}
\end{center}
\end{ruledtabular}
\end{table*}
\subsubsection{CaLi$^+$}
\begin{figure}[ht]
\includegraphics[clip,width=\columnwidth]{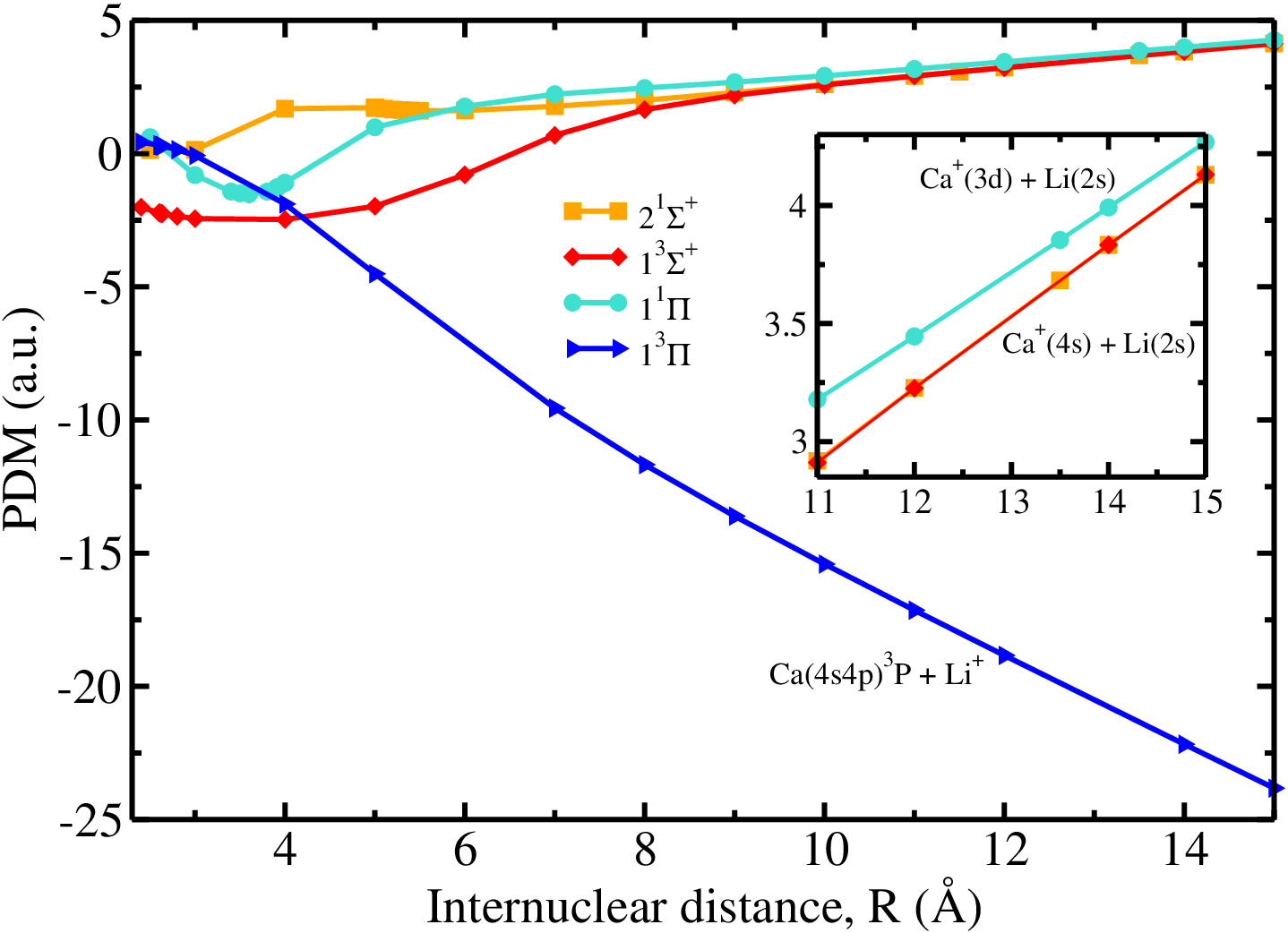}
\caption{\label{fig12}(colour online) Permanent dipole moment curves for the excited states of CaLi$^+$ using EOM-CCSD/QZ method.}
\end{figure}

\begin{figure}[ht]
\includegraphics[clip,width=\columnwidth]{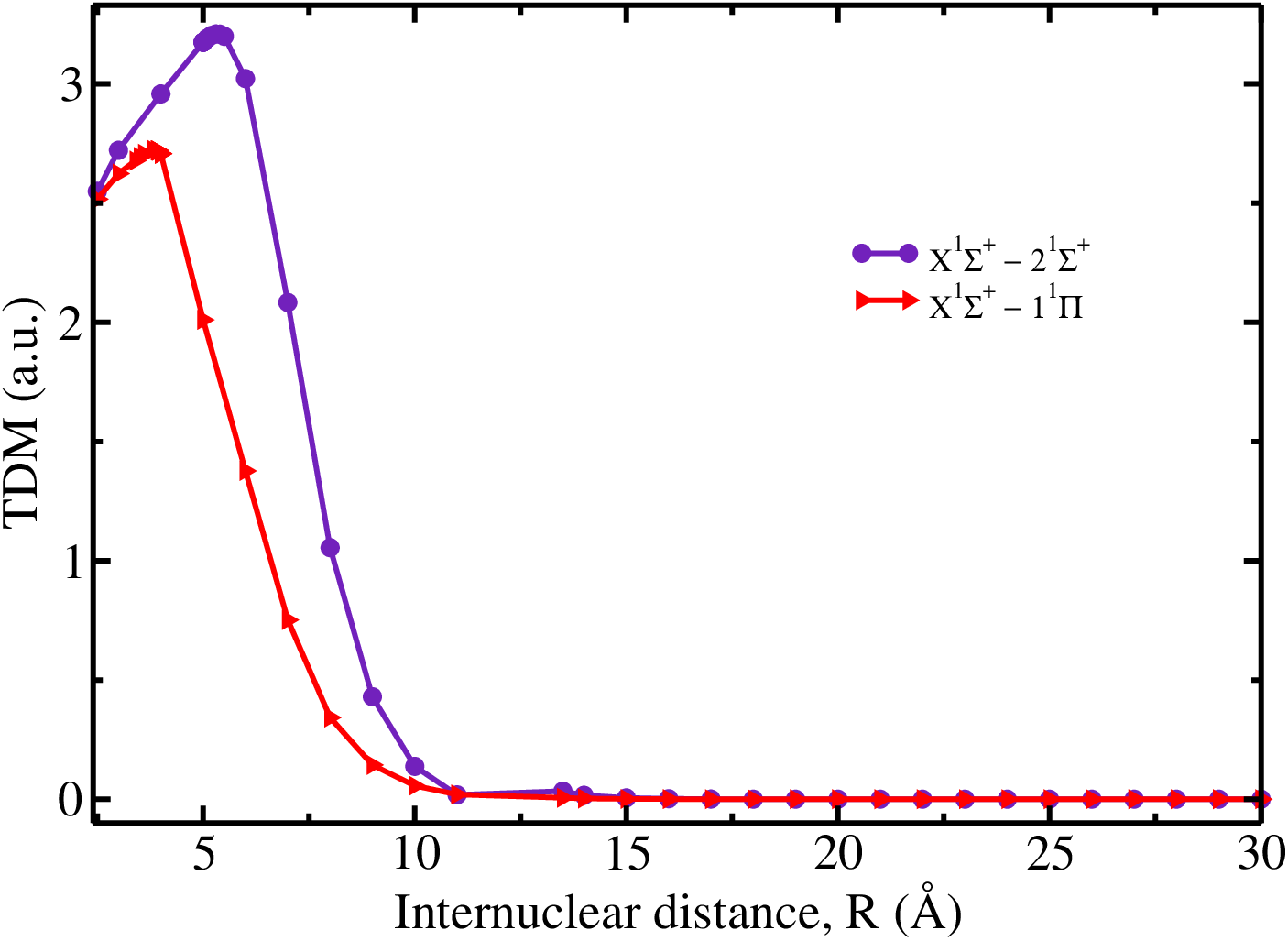}
\caption{\label{fig13}(colour online) TDM curves for the transitions from the electronic ground state to the singlet excited states of CaLi$^+$ using EOM-CCSD/QZ method.}
\end{figure}

\begin{figure}[ht]
\includegraphics[clip,width=\columnwidth]{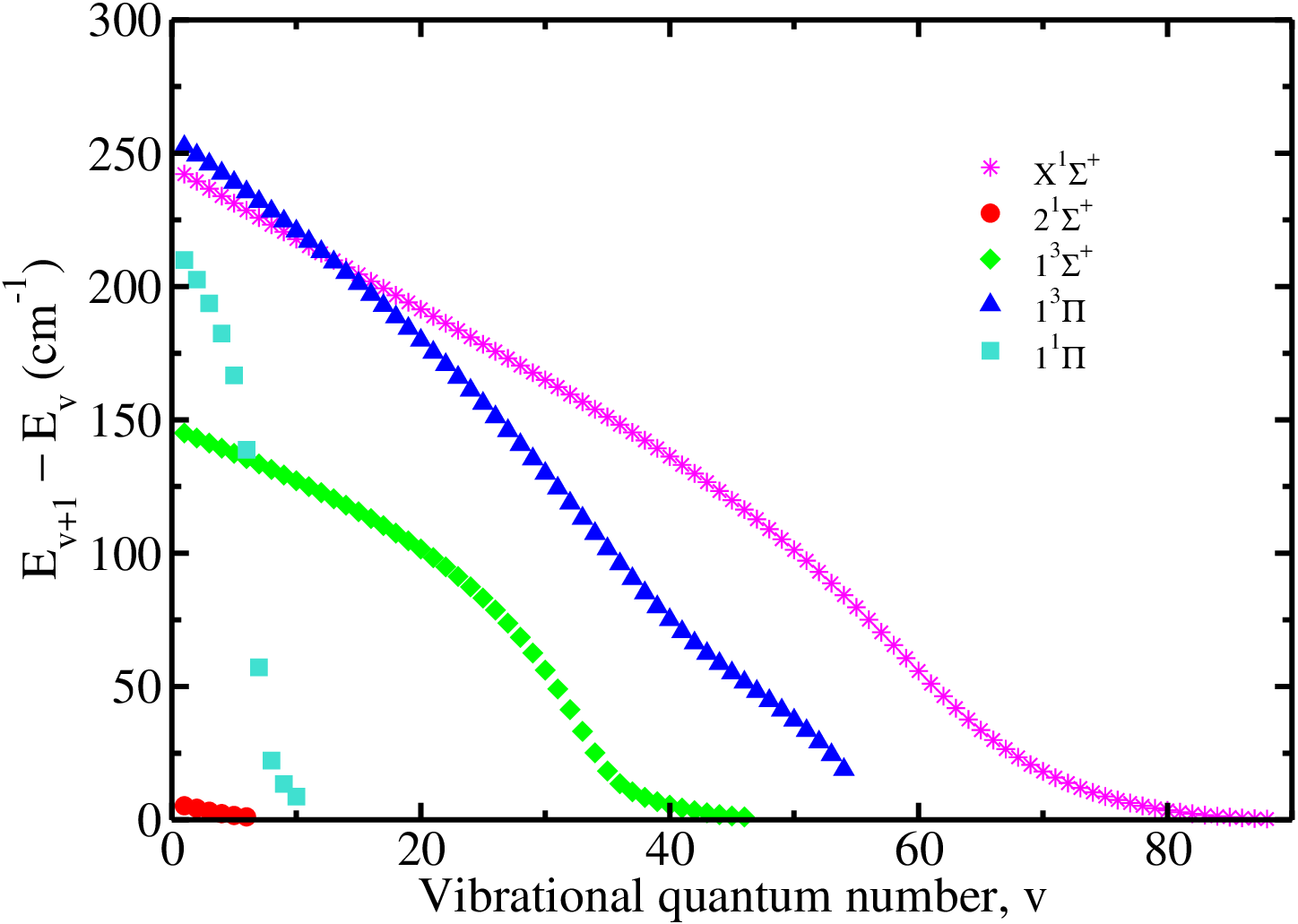}
\caption{\label{fig14}(colour online) Energy spacing between the adjacent vibrational levels in the electronic ground- and excited states of CaLi$^+$.}
\end{figure}

The PECs and PDM curves for the low-lying electronic excited states of CaLi$^+$:  2$^1\Sigma^+$, 1$^3\Sigma^+$, 1$^1\Pi$, and 1$^3\Pi$ are plotted in Figure~\ref{fig11} and Figure~\ref{fig12}, respectively. The derived spectroscopic constants together with the available results in the literature are reported in Table~\ref{table-VI}. The values of equilibrium bond length and harmonic frequency reported in the present work for triplet states are in reasonable agreement with the available results reported in Ref.~\cite{Habli}. But their results for $D_e$ are different from ours by $2195.52$ cm$^{-1}$ for 1$^3\Sigma^+$ and $1361.06$ cm$^{-1}$ for 1$^3\Pi$ state. The reasons for this discrepancy are not different from those discussed earlier for the case of MgLi$^+$.  Although the nature of our PEC for the 2$^1\Sigma^+$ state is very similar to that of Ref.~\cite{Habli}, it is weakly bound in our case and the potential minimum is situated at a distance almost twice of that of Ref.~\cite{Habli}. On the other hand, we have observed one bound minimum situated at a distance of $3.425$ \AA\, for the 1$^1\Pi$ state, and it becomes repulsive after a bond distance of $4.5$ \AA. In contrast, this PEC is reported to be attractive and the state is strongly bound with a potential depth of $6643.47$ cm$^{-1}$  in Ref.~\cite{Habli}. Therefore, our results of the diatomic constants for these singlet states should not be compared directly with that of Ref.~\cite{Habli}. Further theoretical or experimental works are necessary to verify these observations and settle the discrepancies. \\
The atomic states to which these molecular states dissociate into, at the asymptotic distances, are calculated and the compilation is shown in Table~\ref{table-VII}. As it can be seen from this table that both the molecular states 2$^1\Sigma^+$ and 1$^3\Sigma^+$ dissociate into the same set of atomic states, {\itshape viz.}  Ca$^+$\,(4s)\,+\,Li(2s). The excitation energy for these states with respect to the  electronic ground state of neutral CaLi, at the asymptotic limit, is $49875.80$ cm$^{-1}$, and that matches well with the first ionisation energy of Ca, $49305.95$ cm$^{-1}$. The other states 1$^1\Pi$ and 1$^3\Pi$ dissociate into Ca$^+$(3d)\,+\,Li(2s) and Ca\,(4s\,4p)\,$^3$P\,+\,Li$^+$, respectively. The dissociative nature of the molecular states reported in our work agrees with that reported in Ref.~\cite{Habli}. From Figure~\ref{fig12} and Table~\ref{table-VII}, it is clear that those states which possess positive value of dipole moment at large distances, dissociate into Ca\,$^+$\,(4s, {\itshape or} 3d)\,+\,Li\,(2s), whereas, those states with negative value of  dipole moment dissociate into Ca\,(4s$^2$, {\itshape or} 4s4p)\,+\,Li$^+$. This trend is exactly in line with that observed in the case of MgLi$^+$.\\
In order to estimate the effect of diffuse functions on the transition energies, we have used singly-augmented QZ basis sets. The contribution of these additional functions is observed to be less than $0.42$\% at the equilibrium point and $0.063$\% at the asymptotic limit for both the molecular ions. \\
%
%
The TDM curves for the transitions from the electronic ground state to the singlet excited states of CaLi$^+$ are shown in Figure~\ref{fig13}. The maximum value of TDM for X$^1\Sigma^+$ $\rightarrow$ 2$^1\Sigma^+$ and X$^1\Sigma^+$ $\rightarrow$ 1$^1\Pi$ transition is found to be $3.210$ a.u. at $5.3$\,\AA\,and $2.725$ a.u. at $3.8$\,\AA, respectively. At large distance the TDMs drop to zero. \\
We have obtained the maximum vibrational levels to be $7$ for $2^1\Sigma^+$, $47$ for $1^3\Sigma^+$, $55$ for $1^3\Pi$, and $11$ for $1^1\Pi$ state. The corresponding energy spacing between the last two vibrational levels in these states are $1.13$, $1.05$, $8.6$, and $18.8$ cm$^{-1}$, respectively. The relative vibrational energy spacing as a function of $v$ is plotted in Figure~\ref{fig14} and the trends are not uniform because of anharmonic effects. The details of the data related to the vibrational spacings and the rotational constants for the vibrational states are provided in the Supplementary Table S4.\\
%
%
\section{\label{section-4}Summary}
%
In conclusion, we have performed several \emph{ab initio} calculations for the ground- and a few low-lying electronic excited states of MgLi$^+$ and CaLi$^+$. To obtain reliable results for the spectroscopic constants and molecular properties we have used higher-order correlation method such as CCSD(T) and also we have considered all-electron correlations in large basis sets. Our results of diatomic constants and permanent dipole moments are compared with the published results wherever available. The results of the quadrupole moments and various components of dipole polarizabilities of MgLi$^+$ and CaLi$^+$ are being reported  for the first time. The errors arising both from the truncation of correlation effects and finite basis set size effects have also been estimated and the recommended values are highlighted in bold fonts in the tables. Further, the adiabatic effects such as DBOC are estimated for both the ions at the CCSD level. The effect of augmenting the basis sets with diffuse functions on the molecular properties of the ground state, and the transition energies for the excited states is also studied. Using PECs and PDM curves, the vibrational wavefunctions, constants of vibrational spectroscopy, TDMs between different vibrational states, spontaneous and BBR-induced transition rates and hence, the lifetimes of vibrational states are calculated. The lifetime of the  X$^1\Sigma^+$ state in its rovibrational ground state is found to be $2.81\,\mathrm s$ for MgLi$^+$ and $3.19\,\mathrm s$ for CaLi$^+$, at room temperature. The lifetime of vibrational states as a function of the vibrational quantum number is studied and it has been found that the higher vibrational states close to dissociation limit can have lifetimes larger than the lifetime of the rovibrational ground state. It is important to note that the long lifetimes of the higher excited vibrational states are desired for several ultracold experiments. In addition, the TDMs as functions of the internuclear distance are also studied for the transitions from the electronic ground state to the first few singlet excited states for both MgLi$^+$ and CaLi$^+$. We believe that the all-electron \emph{ab initio} results that are presented here may serve as benchmarks for similar calculations in the future and also they may be of interest to the experimental spectroscopists who would consider working on these molecular systems. \\
\begin{center}
{\bf {ACKNOWLEDGMENTS}}
\end{center}
The authors thank Dr. Geetha Gopakumar of Halliburton, Houston, Texas and Prof. Masahiko Hada of TMU, Japan for helpful discussions. R.B. also acknowledges financial support from JST CREST funding for her 1-month stay at TMU. All calculations reported in this work are performed on the computing facility in the Department of Physics, IIT Roorkee, India and Department of Chemistry, TMU Japan.

\end{document}